\documentclass[onecolumn, draftclsnofoot, journal]{IEEEtran}
\usepackage{microtype}
\usepackage{cite}
\usepackage{amsmath,amsfonts,amssymb}
\usepackage{bm}
\usepackage{graphicx}
\usepackage[table,dvipsnames]{xcolor}

\newcommand{\F}{\mathbb{F}}
\renewcommand{\P}{\mathbb{P}}
\newcommand{\Z}{\mathbb{Z}}
\DeclareMathOperator{\wt}{wt}
\DeclareMathOperator{\ev}{ev}

\DeclareMathOperator{\im}{im}

\DeclareMathOperator{\GRS}{GRS}
\DeclareMathOperator{\RS}{RS}
\DeclareMathOperator{\spn}{span}
\DeclareMathOperator{\Mat}{Mat}
\DeclareMathOperator{\supp}{supp}
\DeclareMathOperator{\Dec}{Dec}
\newcommand{\encoding}{\mathrm{enc}}
\newcommand{\security}{\mathrm{sec}}

\newtheorem{theorem}{Theorem}
\newtheorem{lemma}{Lemma}
\newtheorem{proposition}{Proposition}
\newtheorem{corollary}{Corollary}
\newtheorem{definition}{Definition}
\newtheorem{example}{Example}

\newtheorem{remark}{Remark}

\title{General Framework for Linear Secure Distributed Matrix Multiplication with Byzantine Servers}

\author{%
\IEEEauthorblockN{Okko~Makkonen,~\IEEEmembership{Graduate Student Member,~IEEE,}
    and~Camilla~Hollanti,~\IEEEmembership{Member,~IEEE} \\
}%
\IEEEauthorblockA{
    Department of Mathematics and Systems Analysis \\
    Aalto University, Finland \\
    Emails: \{okko.makkonen, camilla.hollanti\}@aalto.fi
}%

\thanks{This work has been supported by the Research Council of Finland under Grant No.\ 336005 and by the Vilho, Yrjö and Kalle Väisälä Foundation of the Finnish Academy of Science and Letters. An earlier version of this paper was presented at the 2022 IEEE Information Theory Workshop~\cite{makkonen2022general}.}%
}

\begin{document}

\maketitle

\begin{abstract}
In this paper, a general framework for linear secure distributed matrix multiplication (SDMM) is introduced. The model allows for a neat treatment of straggling and Byzantine servers via a star product interpretation as well as simplified security proofs. Known properties of star products also immediately yield a lower bound for the recovery threshold as well as an upper bound for the number of colluding workers the system can tolerate. Another bound on the recovery threshold is given by the decodability condition, which generalizes a bound for GASP codes. The framework produces many of the known SDMM schemes as special cases, thereby providing unification for the previous literature on the topic. Furthermore, error behavior specific to SDMM is discussed and interleaved codes are proposed as a suitable means for efficient error correction in the proposed model. Analysis of the error correction capability under natural assumptions about the error distribution is also provided, largely based on well\nobreakdash-known results on interleaved codes. Error detection and other error distributions are also discussed.
\end{abstract}

\begin{IEEEkeywords}
Secure distributed matrix multiplication, Reed--Solomon codes, star product codes, interleaved codes, information-theoretic security.
\end{IEEEkeywords}

\section{Introduction}\label{sec:introduction}

\IEEEPARstart{S}{ecure} distributed matrix multiplication (SDMM) has been studied as a way to compute a matrix product using the help of worker servers such that the computation is information-theoretically secure against colluding workers. SDMM was first studied by Chang and Tandon in \cite{chang2018capacity}. Their scheme was improved by D'Oliveira \emph{et al.}\ in \cite{d2020gasp, d2021degree, d2020notes} using GASP codes. Different schemes have also been introduced in \cite{kakar2019capacity, aliasgari2020private, jia2021capacity, lopez2022secure, mital2022secure, kim2019private, yu2020entangled, karpuk2023modular, byrne2023straggler, machado2022root}. Furthermore, different modes of SDMM, such as private, batch, or cooperative SDMM, have been studied in \cite{chang2019upload, jia2021cross, yu2019lagrange, yu2020entangled, chen2021gcsa, zhu2021secure, li2022efficient, li2022private}. The information-theoretic capacity of SDMM has been studied in \cite{chang2018capacity, jia2021capacity, kakar2019capacity, yang2019secure}, but overall capacity results are still scarce. In addition to considering SDMM over finite fields, SDMM has also been utilized over the analog domain (\emph{i.e.}, real or complex numbers) in \cite{makkonen2022analog}.

The workers in an SDMM scheme are thought of as untrustworthy-but-useful, which means that some of them might not work according to the protocol. The main robustness has been against providing security against colluding workers, which share the information they receive and try to infer the contents of the original matrices. Tools from secret sharing have been used to guarantee information-theoretic security against such colluding workers. Additionally, robustness against so\nobreakdash-called straggling workers has been considered. Stragglers are workers that respond slowly or not at all. Such workers cause an undesired straggler effect if the computation time is limited by the slowest worker.

Byzantine workers are workers that return erroneous results either intentionally or as a result of a fault. Such errors can be difficult to detect directly without further analysis. To guarantee the correctness of the matrix product, it is crucial to be able to detect the errors and correct them with minimal overhead in communication and computation. Tools from classical coding theory can be used to correct errors caused by the Byzantine workers and erasures caused by stragglers.

A coded computation scheme that accounts for stragglers and Byzantine workers has been presented in \cite{yu2019lagrange} using so\nobreakdash-called Lagrange coded computation. This scheme considers stragglers as erasures and Byzantine workers as errors in some linear codes. This means that a straggling worker requires one additional worker and a Byzantine worker requires two additional workers. Furthermore, error detection methods have been utilized in \cite{hofmeister2022secure, tang2022adaptive}. In these methods, the user compares the results given by the workers to the correct results by using probabilistic error detection methods.

\subsection{System Model}

\begin{figure}[!t]
    \centering
    \includegraphics[width=0.6\columnwidth]{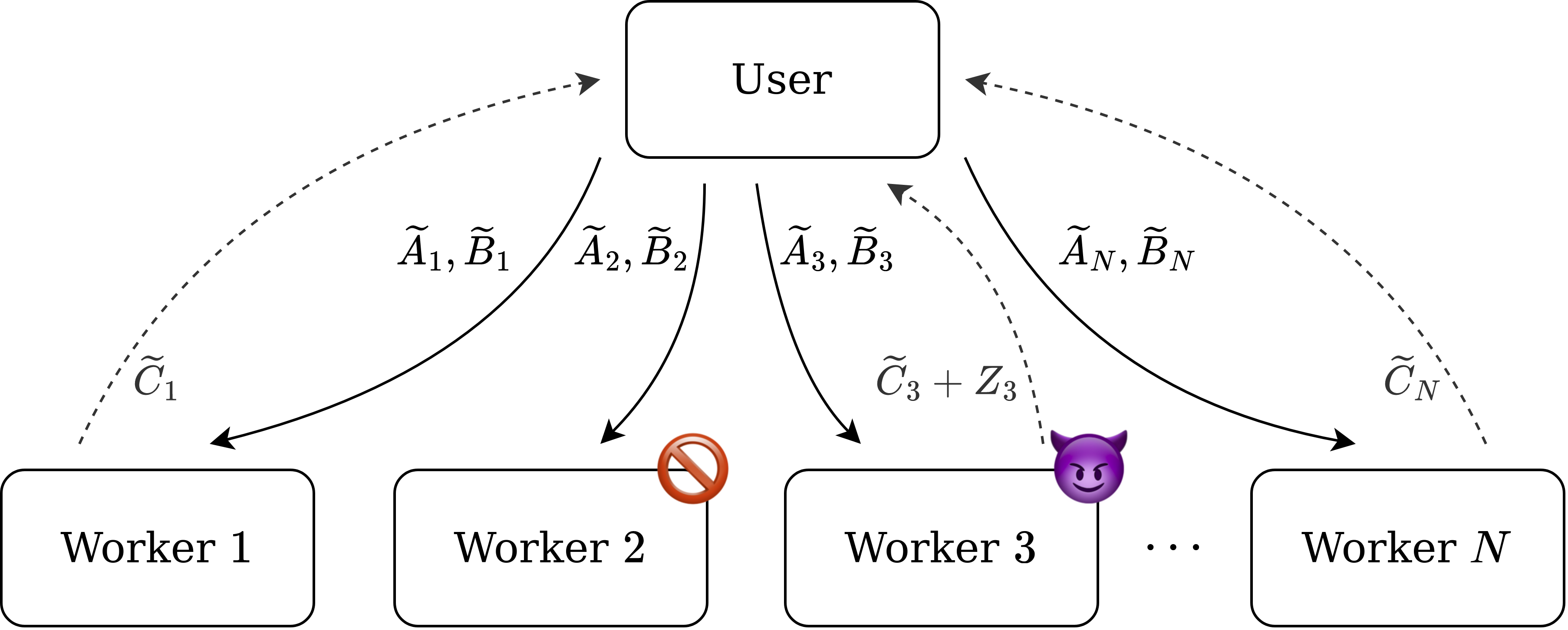}
    \caption{System model of the linear SDMM framework. Worker 2 and 3 are a straggler and a Byzantine worker, respectively.}
    \label{fig:system_model}
\end{figure}

We consider the setting with a user that has two private matrices $A$ and $B$, and access to $N$ workers. The workers receive some encoded pieces $\widetilde{A}_i$, $\widetilde{B}_i$, which are used to compute the response $\widetilde{C}_i$. Some of the users may be stragglers, which means that they do not respond in time. Additionally, some workers may be Byzantine workers, which means that they respond with some erroneous response $\widetilde{C}_i + Z_i$, for some nonzero $Z_i$. These are denoted by workers 2 and 3, respectively, in Figure~\ref{fig:system_model}. The user aims to compute the product $AB$ from the responses.

One of the requirements in SDMM is that the private data contained in the matrices $A$ and $B$ is kept information-theoretically secure from any $X$ colluding workers. The encoded pieces should be made by adding noise to the matrices in such a way that
\begin{equation*}
    I(\bm{A}, \bm{B}; \bm{\widetilde{A}}_\mathcal{X}, \bm{\widetilde{B}}_\mathcal{X}) = 0
\end{equation*}
for all subsets $\mathcal{X}$ of size $X$ of the workers. Here $\bm{\widetilde{A}}_\mathcal{X}$ and $\bm{\widetilde{B}}_\mathcal{X}$ denote the sets of $\bm{\widetilde{A}}_i$ and $\bm{\widetilde{B}}_i$ held by the colluding set $\mathcal{X}$.

There are multiple goals when designing an SDMM scheme, including reducing communication costs, reducing computation time, or increasing robustness against straggling or Byzantine workers. It is a matter of implementation to decide which of these goals to prioritize.

\subsection{Contributions}\label{sec:contributions}

As the main contribution, this paper introduces a general framework for linear SDMM schemes that can be used to construct many SDMM schemes from the literature in a unified way. We show a strong connection between star product codes and SDMM schemes and relate the properties of the associated codes to the security of the schemes as well as to the recovery threshold and collusion tolerance. Previously, star product codes have been successfully utilized in private information retrieval (PIR) \cite{freij2017private}. Using existing results for star product codes, we give new lower bounds for the recovery threshold of linear SDMM schemes in Theorem~\ref{thm:recovery_threshold_stragglers} and Theorem \ref{thm:mds_security_recovery_threshold}. Using these bounds we show that the secure MatDot code presented in \cite{aliasgari2020private} and the SDMM scheme based on the DFT presented in \cite{mital2022secure} are optimal concerning the recovery threshold under some mild assumptions. These bounds are now possible due to the general framework that encompasses many interesting cases, going way beyond the special cases found in the literature. Most previous schemes are based on polynomial evaluation codes, while our framework works for \emph{all} linear codes including algebraic geometry codes. Furthermore, we present a bounded-distance decoding strategy utilizing interleaved codes, which provides robustness against straggling and Byzantine workers. Finally, we analyze the error-correcting capabilities of the proposed strategy under some natural assumptions about the error distributions.

\subsection{Organization}\label{sec:organization}

The organization of this paper is as follows. In Section~\ref{sec:preliminaries} we give some preliminaries on star product codes, and interleaved codes, and introduce the so-called matrix codes. In Section~\ref{sec:examples_of_SDMM_schemes} we give examples of SDMM schemes from the literature. In Section~\ref{sec:general_linear_SDMM_framework} we present our linear SDMM framework and define the decodability and security of such schemes. Additionally, we connect the properties of the scheme with some coding-theoretic notions, which showcases the usefulness of using coding theory to study SDMM. In Section~\ref{sec:security_linear_SDMM} we show a condition for the security of linear SDMM schemes based on the coding-theoretic properties of the scheme. In Section~\ref{sec:linear_SDMM_bounds} we give some fundamental bounds on the recovery threshold of linear SDMM schemes. In particular, we focus on linear SDMM schemes coming from maximum distance separable (MDS) codes. In Section~\ref{sec:constructing_linear_SDMM_schemes} we give examples of linear SDMM schemes based on the SDMM schemes in the literature. In Section~\ref{sec:error_correction_SDMM} we show how interleaved codes and collaborative decoding can be used to treat Byzantine workers in linear SDMM schemes.

\section{Preliminaries}\label{sec:preliminaries}

We write $[n] = \{1, \dots, n\}$. We consider scalars, vectors, and matrices over a finite field $\F_q$ with $q$ elements. The group of units of $\F_q$ is denoted by $\F_q^\times = \F_q \setminus \{0\}$. Vectors in $\F_q^n$ are considered to be row vectors. If $G$ is a matrix, then $G^{\leq m}$ and $G^{> m}$ denote the submatrices with the first $m$ rows and the rest of the rows, respectively. Furthermore, if $\mathcal{I}$ is a set of indices, then $G_\mathcal{I}$ is the submatrix of $G$ with the columns indexed by $\mathcal{I}$. We denote random variables with bold symbols, \emph{i.e.}, the random variable corresponding to $A$ will be denoted by $\bm{A}$.

Throughout, we consider linear codes, \emph{i.e.}, linear subspaces of $\F_q^n$. We denote the dual of a linear code $\mathcal{C}$ by $\mathcal{C}^\perp$. The \emph{support} of a linear code $\mathcal{C} \subseteq \F_q^n$ is defined as $\supp(\mathcal{C}) = \bigcup_{c \in \mathcal{C}} \supp(c)$, where $\supp(c) = \{i \in [n] \mid c_i \neq 0 \}$. We say that $\mathcal{C}$ is of \emph{full-support} if $\supp(\mathcal{C}) = [n]$. A linear code $\mathcal{C}$ is said to be \emph{maximum distance separable (MDS)} if it has minimum distance $d_\mathcal{C} = n - \dim \mathcal{C} + 1$.

\subsection{Star Product Codes}\label{sec:star_product_codes}

The star product is a way of combining two linear codes to form a new linear code. Such a construction has been used in, \emph{e.g.}, code-based cryptography and multiparty computation. A good survey on star products is given in \cite{randriambololona2013upper}.

\begin{definition}[Star product code]
Let $\mathcal{C}$ and $\mathcal{D}$ be linear codes of length $n$ over $\F_q$. The star product of these codes is defined as
\begin{equation*}
    \mathcal{C} \star \mathcal{D} = \spn \{ c \star d \mid c \in \mathcal{C}, d \in \mathcal{D} \},
\end{equation*}
where $(c_1, \dots, c_n) \star (d_1, \dots, d_n) = (c_1d_1, \dots, c_nd_n)$.
\end{definition}

Notice that the star product of codes is defined as the linear span of the elementwise products of codewords. The span is taken so that the resulting code is linear. While the parameters of a star product code are not known in general, we have a Singleton type bound for the minimum distance of a star product of linear codes.

\begin{proposition}[Product Singleton bound \cite{randriambololona2013upper}]\label{prop:product_singleton_bound}
The star product code $\mathcal{C} \star \mathcal{D}$ has minimum distance
\begin{equation*}
    d_{\mathcal{C} \star \mathcal{D}} \leq \max\{ 1, n - (\dim \mathcal{C} +  \dim \mathcal{D}) + 2 \}
\end{equation*}
when $\mathcal{C}$ and $\mathcal{D}$ are linear codes of length $n$.
\end{proposition}

A bound for the dimension of a star product code is given by the following result from \cite{mirandola2015critical}. 

\begin{proposition}\label{prop:star_product_dimension_mds}
Let $\mathcal{C}, \mathcal{D}$ be full-support codes of length $n$. If at least one of the codes is MDS, then
\begin{equation*}
    \dim \mathcal{C} \star \mathcal{D} \geq \min \{ n, \dim \mathcal{C} + \dim \mathcal{D} - 1 \}.
\end{equation*}
\end{proposition}

\subsection{Algebraic Geometry Codes}

In this section, we present some basic notation and concepts on algebraic geometry codes and Reed--Solomon codes. Algebraic geometry codes are linear codes coming from projective smooth irreducible algebraic curves and their associated algebraic function fields. These concepts are included for the interested reader as they are needed for Section~\ref{sec:constructing_linear_SDMM_schemes} but are not needed for the rest of the paper. We follow the presentation in \cite{couvreur2017cryptanalysis} and \cite{stichtenoth2009algebraic}.

Let $F$ be an algebraic function field over $\F_q$ of genus $g$, and $\P_F$ the set of places of $F$. A \emph{divisor} of $F$ is the formal sum
\begin{equation*}
    D = \sum_{P \in \P_F} n_P P,
\end{equation*}
where $n_P \in \Z$ and $n_P \neq 0$ for finitely many $P \in \P_F$. We write $\supp(D) = \{ P \in \P_F \colon n_P \neq 0 \}$ and $\deg D = \sum_{P \in \P_F} n_P \deg P$. We define $D \geq 0$ if $n_P \geq 0$ for all $P \in \P_F$. The principal divisor of $z \in F \setminus \{0\}$ is
\begin{equation*}
    (z) = \sum_{P \in \P_F} v_P(z)P,
\end{equation*}
where $v_P(z)$ is the valuation of $z$ at $P$. The Riemann--Roch space of a divisor $D$ is
\begin{equation*}
    \mathcal{L}(D) = \{ z \in F \setminus \{0\} \colon (z) + D \geq 0 \} \cup \{0\}.
\end{equation*}
This space is a vector space of finite dimension, denoted by $\ell(D)$. Let $\mathcal{P} = \{P_1, \dots, P_n\}$ be a set of distinct rational places. Assume that $\supp(D) \cap \mathcal{P} = \varnothing$. We define the linear map $\ev_\mathcal{P} \colon \mathcal{L}(D) \to \F_q^n$ by
\begin{equation*}
    \ev_\mathcal{P}(z) = (z(P_1), \dots, z(P_n)).
\end{equation*}
The \emph{algebraic geometry code} of places $\mathcal{P}$ and divisor $D$ is
\begin{equation*}
    \mathcal{C}_\mathcal{L}(\mathcal{P}, D) = \ev_\mathcal{P}(\mathcal{L}(D)).
\end{equation*}

We may consider the star product of algebraic geometry codes. From the definition, it is clear that
\begin{equation*}
    \mathcal{C}_\mathcal{L}(\mathcal{P}, D_1) \star \mathcal{C}_\mathcal{L}(\mathcal{P}, D_2) \subseteq \mathcal{C}_\mathcal{L}(\mathcal{P}, D_1 + D_2).
\end{equation*}
Furthermore, if $\deg D_1 \geq 2g + 1$ and $\deg D_2 \geq 2g$, then the above holds with equality \cite{couvreur2017cryptanalysis}.

As a special case, we consider the rational function field $\F_q(x)$. Let $P_\infty$ be the pole of $x$, and let $\mathcal{P} = \{P_1, \dots, P_n\}$ be a set of rational places not containing $P_\infty$. We define the \emph{Reed--Solomon code} as $\mathcal{C}_\mathcal{L}(\mathcal{P}, D)$, where $D = (k-1)P_\infty$ for $k \leq n$. The function $x^i$ is in $\mathcal{L}(D)$ if and only if $(x^i) + D \geq 0$, \emph{i.e.}, if $0 \leq i \leq k - 1$. Therefore, $\mathcal{L}(D) = \{ f(x) \in \F_q[x] \colon \deg f(x) < k \} = \F_q[x]^{<k}$. This leads to the representation
\begin{equation*}
    \RS_k(\alpha) = \{ (f(\alpha_1), \dots, f(\alpha_n)) \mid f(x) \in \F_q[x]^{<k} \},
\end{equation*}
where $P_i = P_{x - \alpha_i}$. It is well-known that $\RS_k(\alpha)$ is an $[n, k]$ MDS code. Furthermore, we define the generalized Reed--Solomon codes as $\GRS_k(\alpha, \nu) = \nu \star \RS_k(\alpha)$ for some vector $\nu \in (\F_q^\times)^n$. As $F$ has genus $g = 0$, we may use the above to get
\begin{equation*}
    \RS_{k_1}(\alpha) \star \RS_{k_2}(\alpha) = \RS_{\min\{n, k_1 + k_2 - 1\}}(\alpha).
\end{equation*}
We notice that the Reed--Solomon codes satisfy the inequalities of Proposition~\ref{prop:product_singleton_bound} and Proposition~\ref{prop:star_product_dimension_mds} with equality.

\subsection{Interleaved Codes}\label{sec:interleaved_codes}

Interleaved codes have been used to correct burst errors in a stream of codewords in many applications. Burst errors are errors where multiple consecutive symbols are affected instead of single symbol errors distributed arbitrarily. These concepts are needed for Section~\ref{sec:error_correction_SDMM}.

\begin{definition}[Homogeneous interleaved codes]
Let $\mathcal{C}$ be a linear code over the field $\F_q$. Then the $\ell$\nobreakdash-interleaved code of $\mathcal{C}$ is the code
\begin{equation*}
    \mathcal{IC}^{(\ell)} = \left\{ \begin{pmatrix}c_1 \\ \vdots \\ c_\ell \end{pmatrix} : c_i \in \mathcal{C}~\forall i \in [\ell] \right\}.
\end{equation*}
\end{definition}

The codewords in an interleaved code are matrices, where each row is a codeword in the code $\mathcal{C}$. Instead of the Hamming weight as the measure of the size of an error, the column weight is used. The column weight of a matrix is defined to be the number of nonzero columns.

When many codewords need to be transmitted, they can be sent such that the first symbol of each codeword is sent, then the second symbol of each codeword, and so on. If a burst error occurs, then multiple codewords are affected, but only a small number of symbols are affected in any particular codeword. This transforms the burst error into single symbol errors in the individual codewords, which means that regular error correction algorithms can be used to correct up to half the minimum distance of errors.

Even more efficient error correction algorithms can be performed for interleaved codes by considering collaborative decoding, where all of the codewords in the interleaved code are considered at the same time. This is advantageous since the error locations in each of the codewords are the same. Collaborative decoding algorithms have been studied in \cite{krachkovsky2003reed, schmidt2009collaborative} and more recently in \cite{holzbaur2021success}. Collaborative decoding algorithms can achieve beyond half the minimum distance decoding by correcting the errors as a system of simultaneous equations.

\subsection{Matrix Codes}\label{sec:matrix_codes}

In this section, we will define matrix codes, which will allow us to consider linear codes whose symbols are matrices of some specified size over the field instead of scalars. This notion can be used to study the algebraic structure of SDMM.

\begin{definition}[Matrix code]\label{def:matrix_code}
Let $\mathcal{C}$ be a linear code of length $n$ over $\F_q$. Then the \emph{$t \times s$ matrix code} of $\mathcal{C}$ is
\begin{equation*}
    \Mat_{t \times s}(\mathcal{C}) = \{ (C_1, \dots, C_n) : C_i \in \F_q^{t \times s}, C^{\alpha\beta} \in \mathcal{C} \}.
\end{equation*}
Here $C^{\alpha\beta} = (C_1^{\alpha\beta}, \dots, C_n^{\alpha\beta})$ is the vector obtained by taking the entry indexed by $(\alpha, \beta) \in [t] \times [s]$ in each of the matrices $C_i$, for $i \in [n]$. Such a code is a linear code in the ambient space $\Mat_{t \times s}(\F_q)^n$.
\end{definition}

We consider the weight of these matrix tuples as the number of nonzero matrices. These objects can be thought of as matrices over the code $\mathcal{C}$, which motivates the notation. Our definition is essentially the same as homogeneous $ts$\nobreakdash-interleaved codes since the matrices contain $ts$ entries. However, this representation leads to some nice multiplicative properties coming from the multiplication of matrices. We define the star product of two such tuples as
\begin{equation*}
    C \star D = (C_1 D_1, \dots, C_n D_n)
\end{equation*}
whenever $C \in \Mat_{t \times s}(\mathcal{C})$ and $D \in \Mat_{s \times r}(\mathcal{D})$. Similarly, we define the star product of the associated spaces by
\begin{align*}
    &\Mat_{t \times s}(\mathcal{C}) \star \Mat_{s \times r}(\mathcal{D}) \\
    &= \spn \{ C \star D \mid C \in \Mat_{t \times s}(\mathcal{C}), D \in \Mat_{s \times r}(\mathcal{D}) \}.
\end{align*}

The following lemma will show that the star product of matrix codes is the matrix code of the star product.

\begin{lemma}\label{lem:matrix_code_star_product}
Let $\mathcal{C}$ and $\mathcal{D}$ be linear codes of length $n$. Then
\begin{equation*}
    \Mat_{t \times s}(\mathcal{C}) \star \Mat_{s \times r}(\mathcal{D}) = \Mat_{t \times r}(\mathcal{C} \star \mathcal{D}).
\end{equation*}
\end{lemma}

\begin{IEEEproof}
Let $\alpha \in [t]$ and $\gamma \in [r]$. By definition of matrix multiplication,
\begin{equation*}
    (C \star D)^{\alpha\gamma}_i = \sum_{\beta=1}^s C_i^{\alpha\beta} D_i^{\beta\gamma}.
\end{equation*}
Therefore, by linearity,
\begin{equation*}
    (C \star D)^{\alpha\gamma} = \sum_{\beta=1}^s C^{\alpha\beta} \star D^{\beta\gamma} \in \mathcal{C} \star \mathcal{D},
\end{equation*}
since $C^{\alpha\beta} \in \mathcal{C}$ and $D^{\beta\gamma} \in \mathcal{D}$. Hence, $C \star D \in \Mat_{t \times r}(\mathcal{C} \star \mathcal{D})$. By linearity of $\Mat_{t \times r}(\mathcal{C} \star \mathcal{D})$, we get that
\begin{equation*}
    \Mat_{t \times s}(\mathcal{C}) \star \Mat_{s \times r}(\mathcal{D}) \subseteq \Mat_{t \times r}(\mathcal{C} \star \mathcal{D}).
\end{equation*}

Fix indices $\alpha \in [t]$ and $\gamma \in [r]$, and codewords $c \in \mathcal{C}$ and $d \in \mathcal{D}$. Let $\beta \in [s]$ and define $C \in \Mat_{t \times s}(\mathcal{C})$ by setting the entries of $C_i$ to be zeros except $C_i^{\alpha\beta} = c_i$. Furthermore, define $D \in \Mat_{s \times r}(\mathcal{D})$ by setting the entries of $D_i$ to be zeros except $D_i^{\beta\gamma} = d_i$. Then,
\begin{equation*}
    (C \star D)_i^{\alpha\gamma} = (C_1D_1)^{\alpha\gamma} = c_id_i
\end{equation*}
so $(C \star D)^{\alpha\gamma} = c \star d$ and the other entries of $C \star D$ are zero vectors. By taking linear combinations of such products we can achieve all codewords in $\Mat_{t \times r}(\mathcal{C} \star \mathcal{D})$, since each entry of such matrices can be represented as a sum of simple star products of the form $c \star d$.
\end{IEEEproof}

We will write just $\Mat(\mathcal{C})$ if the dimensions are clear from context.

\subsection{Examples of SDMM Schemes}\label{sec:examples_of_SDMM_schemes}

In this section, we recall some examples of SDMM schemes by adopting the presentation typically used in the literature. Later, we will show how these schemes arise as special cases from the general framework proposed in this paper.

The goal is to compute the matrix product of the matrices $A \in \F_q^{t \times s}$ and $B \in \F_q^{s \times r}$ using a total of $N$ workers while protecting against any $X$ colluding workers. Furthermore, we denote by $S$ the number of stragglers and by $E$ the number of Byzantine workers. The \emph{recovery threshold} is defined as the number of responses from workers that are required to decode the intended product. In particular, the recovery threshold is the minimal integer $R$ such that \emph{any} $R$ responses are enough to recover the product, but in some cases, fewer than $R$ responses may suffice.

The schemes are based on different matrix partitioning techniques.
The most general matrix partitioning is the \emph{grid partitioning}, which partitions the matrices to $mp$ and $np$ pieces such that
\begin{equation*}
    A = 
    \begin{pmatrix}
        A_{11} & \cdots & A_{1p} \\
        \vdots & \ddots & \vdots \\
        A_{m1} & \cdots & A_{mp}
    \end{pmatrix},
    \quad B = 
    \begin{pmatrix}
        B_{11} & \cdots & B_{1n} \\
        \vdots & \ddots & \vdots \\
        B_{p1} & \cdots & B_{pn}
    \end{pmatrix}.
\end{equation*}
These pieces are obtained by splitting the matrices evenly into the smaller submatrices. The product of these matrices can then be expressed as
\begin{equation*}
    AB = \begin{pmatrix}
        C_{11} & \cdots & C_{1n} \\
        \vdots & \ddots & \vdots \\
        C_{m1} & \cdots & C_{mn}
    \end{pmatrix},
\end{equation*}
where $C_{ik} = \sum_{j=1}^p A_{ij} B_{jk}$. Special cases of this include the \emph{inner product partitioning} (IPP) and \emph{outer product partitioning} (OPP). In IPP the matrices are partitioned into $p$ pieces such that
\begin{equation*}
    A = \begin{pmatrix} A_1 & \cdots & A_p \end{pmatrix}, \quad B = \begin{pmatrix} B_1 \\ \vdots \\ B_p \end{pmatrix}.
\end{equation*}
Then the product can be expressed as $AB = \sum_{j=1}^p A_j B_j$. In OPP the matrices are partitioned into $m$ and $n$ pieces, respectively, such that
\begin{equation*}
    A = \begin{pmatrix} A_1 \\ \vdots \\ A_m \end{pmatrix}, \quad B = \begin{pmatrix} B_1 & \cdots & B_n \end{pmatrix}.
\end{equation*}
Then the product can be expressed as
\begin{equation*}
    AB = \begin{pmatrix}
    A_1B_1 & \cdots & A_1B_n \\
    \vdots & \ddots & \vdots \\
    A_mB_1 & \cdots & A_mB_n
    \end{pmatrix}.
\end{equation*}

In the next three examples, we will present some well-known examples from the literature.

\begin{example}[Secure MatDot \cite{aliasgari2020private}]\label{ex:secure_matdot}
The secure MatDot scheme uses the inner product partitioning to split the matrices into $p$ pieces. Define the polynomials
\begin{align*}
    f(x) &= \sum_{j=1}^p A_j x^{j-1} + \sum_{k=1}^X R_k x^{p+k-1}, \\
    g(x) &= \sum_{j'=1}^p B_{j'} x^{p-j'} + \sum_{k'=1}^X S_{k'}x^{p+k'-1},
\end{align*}
where $R_1, \dots, R_X$ and $S_1, \dots, S_X$ are matrices of appropriate size that are chosen uniformly at random over $\F_q$. Let $\alpha_1, \dots, \alpha_N \in \F_q^\times$ be distinct nonzero points and evaluate the polynomials $f(x)$ and $g(x)$ at these points to get the encoded matrices
\begin{equation*}
    \widetilde{A}_i = f(\alpha_i), \quad \widetilde{B}_i = g(\alpha_i).
\end{equation*}
These encoded matrices can be sent to each worker node. The workers compute the matrix products $\widetilde{C}_i = \widetilde{A}_i \widetilde{B}_i$ and return these to the user. The user receives evaluations of the polynomial $h(x) = f(x)g(x)$ from each worker. Using the definition of $f(x)$ and $g(x)$ we can write out the coefficients of $h(x)$ as
\begin{equation*}
    h(x) = \sum_{j=1}^p \sum_{j'=1}^p A_jB_{j'} x^{p+j-j'-1} + (\text{terms of degree $\geq p$}).
\end{equation*}
The degree of $h(x)$ is at most $2p + 2X - 2$. Furthermore, the coefficient of the term $x^{p-1}$ is exactly the product $AB$, which we wish to recover. Using polynomial interpolation we can compute the required coefficient, given that we have at least $2p + 2X - 1$ evaluations. Therefore, the recovery threshold of the secure MatDot code is $R = 2p + 2X - 1$.
\end{example}

\begin{example}[GASP \cite{d2020gasp}]\label{ex:gasp}
Similar to Example~\ref{ex:secure_matdot}, this scheme is also based on polynomial evaluation, but the choice of the polynomials and the evaluation points is more involved. Additionally, the matrices are partitioned according to the outer product partitioning. The following example will give an idea of the general construction described in \cite{d2020gasp, d2021degree}.

The matrices $A \in \F_q^{t \times s}$ and $B \in \F_q^{s \times r}$ are split into $m = n = 3$ submatrices with the outer product partitioning. We wish to protect against $X = 2$ colluding workers. Define the polynomials
\begin{align*}
    f(x) &= A_1 + A_2x + A_3x^2 + R_1x^9 + R_2x^{12}, \\
    g(x) &= B_1 + B_2x^3 + B_3x^6 + S_1x^9 + S_2x^{10},
\end{align*}
where $R_1, R_2, S_1, S_2$ are matrices of appropriate size that are chosen uniformly at random over $\F_q$. The exponents are chosen carefully so that the total number of workers needed is as low as possible. Let $\alpha_1, \dots, \alpha_N \in \F_q^\times$ be distinct nonzero points and evaluate the polynomials $f(x)$ and $g(x)$ at these points to get the encoded matrices
\begin{equation*}
    \widetilde{A}_i = f(\alpha_i), \quad \widetilde{B}_i = g(\alpha_i).
\end{equation*}
These encoded matrices can be sent to each worker node. The workers compute the matrix products $\widetilde{C}_i = \widetilde{A}_i \widetilde{B}_i$ and send these to the user. The user receives evaluations of the polynomial $h(x) = f(x)g(x)$ from each worker. Using the definition of $f(x)$ and $g(x)$ we can write out the coefficients of $h(x)$ as
\begin{align*}
    h(x) &= A_1B_1 + A_2B_1x + A_3B_1x^2 + A_1B_2x^3 + A_2B_2x^4 \\ 
    &+ A_2B_3x^5 + A_1B_3x^6 + A_2B_3x^7 + A_3B_3x^8 \\
    &+ (\text{terms of degree $\geq 9$}).
\end{align*}
We notice that the coefficients of the first $9$ terms are exactly the submatrices we wish to recover. We need 18 responses from the workers, since $h(x)$ has $18$ nonzero coefficients, provided that the corresponding linear equations are solvable. In this case, the recovery threshold is $R = 18$.

The general choice of the exponents in the polynomials $f(x)$ and $g(x)$ is explained in \cite{d2021degree}. A so-called degree table is used to analyze the recovery threshold of the scheme. Furthermore, the choice of the evaluation points is not as simple as with the secure MatDot code, but it was shown that a suitable choice can be made in a large enough field \cite{d2020gasp}.
\end{example}

\begin{example}[SDMM based on DFT \cite{mital2022secure}]\label{ex:sdmm_based_on_dft}
In the SDMM scheme based on the discrete Fourier transform, the matrices are split into $p = N - 2X$ pieces with the inner product partitioning. Define the functions
\begin{align*}
    f(x) &= \sum_{j=1}^p A_j x^{j-1} + \sum_{k=1}^X R_k x^{p + k - 1}, \\
    g(x) &= \sum_{j'=1}^p B_{j'} x^{-j'+1} + \sum_{k'=1}^X S_{k'} x^{-p-X-k'+1},
\end{align*}
where $R_1, \dots, R_X$ and $S_1, \dots, S_X$ are matrices of appropriate size that are chosen uniformly at random over $\F_q$. Let $\zeta \in \F_q^\times$ be a primitive $N$th root of unity. The functions $f(x)$ and $g(x)$ are evaluated at the points $1, \zeta, \zeta^2, \dots, \zeta^{N-1}$ and the results are sent to the workers such that worker $i \in [N]$ receives the encoded matrices
\begin{equation*}
    \widetilde{A}_i = f(\zeta^{i-1}), \quad \widetilde{B}_i = g(\zeta^{i-1}).
\end{equation*}
The workers compute the matrix products of the encoded matrices and return the results $\widetilde{C}_i = \widetilde{A}_i \widetilde{B}_i$. The user receives evaluations of the function
\begin{equation*}
    h(x) = f(x)g(x) = \sum_{j=1}^p A_j B_j + (\text{non-constant terms}).
\end{equation*}
The other terms have degree in $[-N + 1, N - 1]$, which means that the average of the responses equals the constant term, since $\sum_{i=1}^N \zeta^s = 0$ for $N \nmid s$. Hence, the product $AB$ can be computed as the average of all the responses. This means that no stragglers can be tolerated since all of the responses are needed. Furthermore, the field has to be such that the appropriate $N$th root of unity exists.
\end{example}

\section{Linear SDMM}\label{sec:linear_SDMM}

Many SDMM schemes in the literature use concepts from coding theory and secret sharing but are usually presented as concrete constructions based on polynomial interpolation. This makes it easy to argue that the schemes compute the desired matrix product, but the comparison of different schemes is difficult. A more general and abstract description can provide simpler comparisons between SDMM schemes, as well as allow for constructions that are not based on any particular SDMM scheme while losing some detail about why each scheme works the way they do. In this section, we present a general linear SDMM framework that can be used to describe the earlier SDMM schemes compactly. This scheme uses the common elements of each of the examples presented in the previous section. Furthermore, we prove a general security result for linear SDMM schemes and give some bounds on the recovery threshold.

\subsection{A General Linear SDMM Framework via Star Products}\label{sec:general_linear_SDMM_framework}

A linear SDMM scheme over the field $\F_q$ can be constructed in general with the following formula. Here $N$ denotes the total number of workers, $X$ the designed security parameter, and $m, p, n$ partitioning parameters.
\begin{itemize}
\item The input matrices $A \in \F_q^{t \times s}$ and $B \in \F_q^{s \times r}$ are split into submatrices $A_1, \dots, A_{mp}$ and $B_1, \dots, B_{np}$ using the grid partitioning and some enumeration of the partitions.

\item Matrices $R_1, \dots, R_X$ and $S_1, \dots, S_X$ are drawn uniformly at random such that the matrices $R_k$ and $S_{k'}$ have the same dimensions as the partitions of $A$ and $B$, respectively.

\item By combining the partitions and the random matrices we get the following tuples of matrices
\begin{align*}
    &(A_1, \dots, A_{mp}, R_1, \dots, R_X), \\
    &(B_1, \dots, B_{np}, S_1, \dots, S_X)
\end{align*}
of length $mp + X$ and $np + X$, respectively. These tuples are encoded using linear codes $\mathcal{C}_A$ and $\mathcal{C}_B$ of length $N$. Let $F$ and $G$ be suitable generator matrices of size \mbox{$(mp + X) \times N$} and $(np + X) \times N$ for $\mathcal{C}_A$ and $\mathcal{C}_B$, respectively. The encoded matrices are then
\begin{align*}
    \widetilde{A} &= (\widetilde{A}_1, \dots, \widetilde{A}_N) = (A_1, \dots, A_{mp}, R_1, \dots, R_X) F, \\
    \widetilde{B} &= (\widetilde{B}_1, \dots, \widetilde{B}_N) = (B_1, \dots, B_{np}, S_1, \dots, S_X) G.
\end{align*}

\item Each worker is sent one component of each vector, \emph{i.e.}, worker $i \in [N]$ receives matrices $\widetilde{A}_i$ and $\widetilde{B}_i$. The worker then computes $\widetilde{A}_i \widetilde{B}_i$ and sends the result to the user. In coding-theoretic terms, this can be interpreted as the star product of the vectors $\widetilde{A}$ and $\widetilde{B}$. Hence, we may write
\begin{equation*}
    \widetilde{C} = \widetilde{A} \star \widetilde{B} = (\widetilde{A}_1 \widetilde{B}_1, \dots, \widetilde{A}_N \widetilde{B}_N).
\end{equation*}

\item The user computes a linear combination of the responses $\widetilde{C}_i$ to obtain the product $AB$. Not all of the responses may be needed, which means that the scheme can tolerate straggling workers.
\end{itemize}

By definition of matrix codes in Definition~\ref{def:matrix_code} we have that
\begin{equation*}
    \widetilde{A} \in \Mat(\mathcal{C}_A), \quad \widetilde{B} \in \Mat(\mathcal{C}_B)
\end{equation*}
since these tuples were obtained by multiplication by the generator matrices. Therefore,
\begin{equation*}
    \widetilde{C} = \widetilde{A} \star \widetilde{B} \in \Mat(\mathcal{C}_A \star \mathcal{C}_B)
\end{equation*}
by Lemma~\ref{lem:matrix_code_star_product}. However, $\widetilde{C}$ does not generally consist of elementary products $c_A \star c_B$ for $c_A \in \mathcal{C}_A$ and $c_B \in \mathcal{C}_B$. As $\widetilde{A}$ can be any element in $\Mat(\mathcal{C}_A)$ and $\widetilde{B}$ can be any element of $\Mat(\mathcal{C}_B)$, we can achieve all elements of $\Mat(\mathcal{C}_A \star \mathcal{C}_B)$ as linear combinations of the responses $\widetilde{C} = \widetilde{A} \star \widetilde{B}$ by Lemma~\ref{lem:matrix_code_star_product}. Hence, the smallest linear code that the responses live in is $\Mat(\mathcal{C}_A \star \mathcal{C}_B)$, even though the responses do not necessarily form a linear subspace.

We will denote the encoding of the matrix and the encoding of the random padding by
\begin{align*}
    A' &= (A_1, \dots, A_{mp})F^{\leq mp}, & R' &= (R_1, \dots, R_X)F^{> mp}, \\
    B' &= (B_1, \dots, B_{np})G^{\leq np}, & S' &= (S_1, \dots, S_X)G^{> np}.
\end{align*}
Then we have that $\widetilde{A} = A' + R'$ and $\widetilde{B} = B' + S'$. This corresponds to the decomposition
\begin{align*}
    \mathcal{C}_A &= \mathcal{C}_A^\encoding + \mathcal{C}_A^\security, \\
    \mathcal{C}_B &= \mathcal{C}_B^\encoding + \mathcal{C}_B^\security,
\end{align*}
where $\mathcal{C}_A^\encoding$ and $\mathcal{C}_B^\encoding$ are generated by $F^{\leq mp}$ and $G^{\leq np}$, respectively, and $\mathcal{C}_A^\security$ and $\mathcal{C}_B^\security$ are generated by $F^{> mp}$ and $G^{> np}$, respectively. These codes denote the encoding of the matrices and the security part, respectively.

Next, we define what the last step of the linear SDMM framework means, \emph{i.e.}, how the linear combinations of the responses give us the product $AB$. The decodability of SDMM schemes has previously been defined by stating that the product $AB$ can be computed using some unknown function. Here we require that the function is linear since we are in the linear SDMM setting.

\begin{definition}\label{def:decodability}
Let $\mathcal{K} \subseteq [N]$. A linear SDMM scheme is \emph{$\mathcal{K}$\nobreakdash-decodable} if there exist matrices $\Lambda^\mathcal{K}_i \in \F_q^{m \times n}$ such that
\begin{equation*}
    AB = \sum_{i \in \mathcal{K}} \Lambda^\mathcal{K}_i \otimes \widetilde{C}_i,
\end{equation*}
for all matrices $A$ and $B$ and all choices of the random matrices $R_k$ and $S_{k'}$. Here, $\otimes$ denotes the Kronecker product. In particular, we say that a linear SDMM scheme is \emph{decodable} if it is $[N]$\nobreakdash-decodable. In this case we write $\Lambda_i = \Lambda_i^{[N]}$.
\end{definition}

Notice that we do not allow $\Lambda_i$ to depend on the random matrices. The reason for this is that the decoding process should not involve expensive computations by the user. The following lemma will show which responses are required for decoding.

\begin{lemma}\label{lem:information_set_decoding}
Consider a decodable linear SDMM scheme and an information set $\mathcal{I} \subseteq [N]$ of $\mathcal{C}_A \star \mathcal{C}_B$. Then the linear SDMM scheme is $\mathcal{I}$\nobreakdash-decodable. 
In particular, the decoding can be done from any $N - D + 1$ responses, where $D$ is the minimum distance of $\mathcal{C}_A \star \mathcal{C}_B$.
\end{lemma}

\begin{IEEEproof}
Let $H$ be a generator matrix for $\mathcal{C}_A \star \mathcal{C}_B$ and $\mathcal{I} \subseteq [N]$ an information set of $\mathcal{C}_A \star \mathcal{C}_B$. Then,
\begin{equation*}
    \widetilde{C} = \widetilde{C}_\mathcal{I} (H_\mathcal{I})^{-1} H,
\end{equation*}
\emph{i.e.}, the whole response can be computed only from the responses from an information set $\mathcal{I}$. In particular, there are coefficients $\lambda^\mathcal{I}_{ij}$ such that
\begin{equation*}
    \widetilde{C}_i = \sum_{j \in \mathcal{I}} \lambda^\mathcal{I}_{ij} \widetilde{C}_j.
\end{equation*}
Thus,
\begin{equation*}
    AB = \sum_{i \in [N]} \Lambda_i \otimes \bigg( \sum_{j \in \mathcal{I}} \lambda^\mathcal{I}_{ij} \widetilde{C}_j \bigg) = \sum_{j \in \mathcal{I}} \underbrace{\bigg( \sum_{i \in [N]} \lambda^\mathcal{I}_{ij} \Lambda_i \bigg)}_{=\Lambda^\mathcal{I}_j} \otimes\,\widetilde{C}_j.
\end{equation*}
Hence, the product $AB$ can be computed from just the responses from an information set.

Let $\mathcal{K} \subseteq [N]$ be such that $|\mathcal{K}| \geq N - D + 1$. Then the projection from $\mathcal{C}_A \star \mathcal{C}_B$ to the coordinates indexed by $\mathcal{K}$ is injective by definition of minimum distance. Hence, $\mathcal{K}$ contains an information set, so the product can be decoded from the responses of $\mathcal{K}$.
\end{IEEEproof}

In addition to being able to decode the result from any $N - D + 1$ responses, there is also a set of $N - D$ indices that do not contain an information set. Therefore, it is natural to define the recovery threshold of a linear SDMM scheme as $R = N - D + 1$. This means that the scheme can tolerate at most $D - 1$ stragglers. If $\mathcal{C}_A \star \mathcal{C}_B$ is an $[N, K, D]$ MDS code, then we have that $R = K$, which is minimal by the Singleton bound.

In \cite{lopez2022secure} the authors show that using their secure MatDot construction it is possible to recover the result from a smaller number of fixed workers. This does not contradict our definition of recovery threshold, since we require that the result can be recovered from any $R$ responses from the workers.

In addition to decodability, we define the security of linear SDMM schemes.

\begin{definition}\label{def:security}
An SDMM scheme is said to be \emph{secure against $X$\nobreakdash-collusion} (or \emph{$X$\nobreakdash-secure}) if
\begin{equation*}
    I(\bm{A}, \bm{B}; \bm{\widetilde{A}}_\mathcal{X}, \bm{\widetilde{B}}_\mathcal{X}) = 0
\end{equation*}
for all $\mathcal{X} \subseteq [N]$, $|\mathcal{X}| \leq X$, and all distributions of $\bm{A}$ and $\bm{B}$.
\end{definition}

The above definition is the same that has previously been considered in the literature with the exception that the distribution of $\bm{A}$ and $\bm{B}$ has not been explicitly mentioned. We require that the scheme is secure for all possible distributions to avoid some uninteresting edge cases. In particular, any SDMM scheme is secure if we only look at distributions such that $H(\bm{A}) = H(\bm{B}) = 0$. In practice, we will work with uniformly distributed $\bm{A}$ and $\bm{B}$, since this maximizes the entropy.

This construction of linear SDMM schemes is quite abstract as it does not provide a general way of constructing new SDMM schemes from any linear codes. However, it provides a robust and general way to study different SDMM schemes and prove general results. The security properties are determined by the codes $\mathcal{C}_A^\security$ and $\mathcal{C}_B^\security$ as the following lemma and Proposition~\ref{prop:linear_SDMM_security} show.

\begin{lemma}\label{lem:not_too_secure}
A decodable linear SDMM scheme is not $\min\{\dim \mathcal{C}_A^\security + 1, \dim \mathcal{C}_B^\security + 1\}$\nobreakdash-secure.
\end{lemma}

\begin{IEEEproof}
Without loss of generality, let us consider an information set $\mathcal{I} \subseteq [N]$ of $\mathcal{C}_A$. Then $|\mathcal{I}| = \dim \mathcal{C}_A$. As the scheme has to be decodable, we must have that $\dim \mathcal{C}_A > \dim \mathcal{C}_A^\security$, since otherwise the encoded pieces would only be determined by randomness. Consider a set $\mathcal{X} \subseteq \mathcal{I}$ such that $|\mathcal{X}| = \dim \mathcal{C}_A^\security + 1$. Thus, the columns of $F^{>mp}_\mathcal{X}$ are linearly dependent, but the columns of $F_\mathcal{X}$ are linearly independent. Therefore,
\begin{align*}
    I(\bm{A}; \bm{\widetilde{A}}_\mathcal{X}) &= H(\bm{\widetilde{A}}_\mathcal{X}) - H(\bm{\widetilde{A}}_\mathcal{X} \mid \bm{A}) \\
    &= H(\bm{\widetilde{A}}_\mathcal{X}) - H(\bm{A}'_\mathcal{X} + \bm{R}'_\mathcal{X} \mid \bm{A}) \\
    &= H(\bm{\widetilde{A}}_\mathcal{X}) - H(\bm{R}'_\mathcal{X}) > 0.
\end{align*}
Here we used the definition of mutual information, the decomposition of $\bm{\widetilde{A}} = \bm{A}' + \bm{R}'$, the fact that $\bm{A}'$ is completely determined by $\bm{A}$, and $\bm{R}'$ is independent of $\bm{A}$. Finally, $\bm{\widetilde{A}}_\mathcal{X}$ is uniformly distributed, but $\bm{R}'_\mathcal{X}$ is not. As $|\mathcal{X}| = \dim \mathcal{C}_A^\security + 1$, the scheme is not secure against $(\dim \mathcal{C}_A^\security+1)$\nobreakdash-collusion.
\end{IEEEproof}

Now, we can show that the linear codes $\mathcal{C}_A$ and $\mathcal{C}_B$ have the expected dimensions.

\begin{proposition}
The codes $\mathcal{C}_A$ and $\mathcal{C}_B$ of a decodable and $X$\nobreakdash-secure linear SDMM scheme have dimensions $mp + X$ and $np + X$, respectively.
\end{proposition}

\begin{IEEEproof}
The generator matrix $F$ has dimensions $(mp + X) \times N$, so we need to show that $F$ has full row rank.

If the $X \times N$ matrix $F^{> mp}$ does not have full row rank, then $\dim \mathcal{C}_A^\security \leq X - 1$ so by Lemma~\ref{lem:not_too_secure} the scheme is not $X$\nobreakdash-secure. Hence, $F^{> mp}$ has full row rank. 

Assume that $F$ does not have full row rank. Then there is a matrix $A$ and random matrices $R_k$ such that
\begin{equation*}
    \widetilde{A} = (A_1, \dots, A_{mp}, R_1, \dots, R_X)F = 0.
\end{equation*}
We must have that $A \neq 0$, since otherwise $F^{> mp}$ would not have full row rank. Let us choose $B$ such that $AB \neq 0$. Then, $\widetilde{C} = \widetilde{A} \star \widetilde{B} = 0$, but from the decodability we get that
\begin{equation*}
    0 \neq AB = \sum_{i \in [N]} \Lambda_i \otimes \widetilde{C}_i = 0.
\end{equation*}
Hence, $F$ has full row rank. A similar argument shows that $G$ has full row rank.
\end{IEEEproof}

We can now write the earlier decomposition as
\begin{align*}
    \mathcal{C}_A &= \mathcal{C}_A^\encoding \oplus \mathcal{C}_A^\security, \\
    \mathcal{C}_B &= \mathcal{C}_B^\encoding \oplus \mathcal{C}_B^\security,
\end{align*}
where $\dim \mathcal{C}_A^\encoding = mp$, $\dim \mathcal{C}_B^\encoding = np$, and $\dim \mathcal{C}_A^\security = \dim \mathcal{C}_B^\security = X$. By projecting to $\supp(\mathcal{C}_A \star \mathcal{C}_B) = \supp(\mathcal{C}_A) \cap \supp(\mathcal{C}_B)$, we may assume that $\mathcal{C}_A$ and $\mathcal{C}_B$ are full-support codes since this does not affect the properties of the star products. Furthermore, $\mathcal{C}_A^\security$ and $\mathcal{C}_B^\security$ must have full support since otherwise there is no randomness added to one of the encoded pieces.

\begin{remark}
The communication costs incurred by the linear SDMM framework can be computed as follows. Here the costs are measured as the number of $\F_q$ symbols. The user needs to upload $N$ matrices of size $\frac{t}{m} \times \frac{s}{p}$ and $N$ matrices of size $\frac{s}{p} \times \frac{r}{n}$ for a total upload cost of $N(\frac{ts}{mp} + \frac{sr}{pn})$. The user needs to download $R$ matrices of size $\frac{t}{m} \times \frac{r}{n}$ for a total download cost of $R\frac{tr}{mn}$. The total communication cost is then $N(\frac{ts}{mp} + \frac{sr}{pn}) + R\frac{tr}{mn}$. As $N$ can be made as small as $R$, given some fixed matrix partitioning $m, p, n$ the communication cost is essentially determined by the recovery threshold $R$ as well as the matrix dimensions $t, s, r$. The parameters $m, n, p$ can be optimized to find a suitable compromise between communication and computation.
\end{remark}

\subsection{Security of Linear SDMM Schemes}\label{sec:security_linear_SDMM}

The security of linear SDMM comes from the fact that the schemes implement a secret sharing scheme such as the one introduced by Shamir in \cite{shamir1979share}. The following proposition is a well-known result in secret sharing and will highlight the usefulness of the linear SDMM framework since the security of the schemes can be proven by checking the properties of the codes $\mathcal{C}_A^\security$ and $\mathcal{C}_B^\security$. A version of this theorem has been stated in, \emph{e.g.} \cite{pieprzyk2003ideal}.
Recall that a matrix is the generator matrix of an MDS code if and only if all of its maximal submatrices are invertible.

\begin{proposition}\label{prop:linear_SDMM_security}
A linear SDMM scheme is $X$\nobreakdash-secure if $\mathcal{C}_A^\security$ and $\mathcal{C}_B^\security$ are MDS codes.
\end{proposition}

\begin{IEEEproof}
Let $\mathcal{X} \subseteq [N]$, $|\mathcal{X}| = X$, be a set of $X$ colluding nodes. Writing the generator matrix $F$ as
\begin{equation*}
    F = \begin{pmatrix}
    F^{\leq mp} \\
    F^{>mp}
    \end{pmatrix}
\end{equation*}
allows us to write the shares the colluding nodes have about the encoded matrix $\widetilde{A}$ as
\begin{equation*}
    \bm{\widetilde{A}}_\mathcal{X} = \underbrace{(\bm{A}_1, \dots, \bm{A}_{mp})F^{\leq mp}_\mathcal{X}}_{=\bm{A}'_\mathcal{X}} + \underbrace{(\bm{R}_1, \dots, \bm{R}_X)F^{> mp}_\mathcal{X}}_{=\bm{R}'_\mathcal{X}}.
\end{equation*}
If $\mathcal{C}_A^\security$ is an MDS code, then any $X \times X$ submatrix of $F^{> mp}$ is invertible. As $(\bm{R}_1, \dots, \bm{R}_X)$ is uniformly distributed, we get that $\bm{R}'_\mathcal{X} = (\bm{R}_1, \dots, \bm{R}_X)F^{> mp}_\mathcal{X}$ is also uniformly distributed. Therefore,
\begin{align*}
    0 \leq I(\bm{A}; \bm{\widetilde{A}}_\mathcal{X}) &= H(\bm{\widetilde{A}}_\mathcal{X}) - H(\bm{\widetilde{A}}_\mathcal{X} \mid \bm{A}) \\
    &= H(\bm{\widetilde{A}}_\mathcal{X}) - H(\bm{A}'_\mathcal{X} + \bm{R}'_\mathcal{X} \mid \bm{A}) \\
    &= H(\bm{\widetilde{A}}_\mathcal{X}) - H(\bm{R}'_\mathcal{X}) \leq 0,
\end{align*}
since a uniform distribution maximizes the entropy. Here we used the fact that $\bm{A}'_\mathcal{X}$ is completely determined by $\bm{A}$. The idea is that the confidential data of $\bm{A}$ is hidden by adding uniformly random noise. A similar argument works for the matrix $B$. Finally, we get that 
\begin{align*}
    0 &\leq I(\bm{A}, \bm{B}; \bm{\widetilde{A}}_\mathcal{X}, \bm{\widetilde{B}}_\mathcal{X}) \\
    &= I(\bm{A}, \bm{B}; \bm{\widetilde{A}}_\mathcal{X}) + I(\bm{A}, \bm{B}; \bm{\widetilde{B}}_\mathcal{X} \mid \bm{\widetilde{A}}_\mathcal{X}) \\
    &\leq I(\bm{A}; \bm{\widetilde{A}}_\mathcal{X}) + I(\bm{B}; \bm{\widetilde{B}}_\mathcal{X}) = 0.
\end{align*}
The inequality follows from $\bm{\widetilde{A}}_\mathcal{X}$ being conditionally independent of $\bm{B}$ given $\bm{A}$, and $\bm{\widetilde{B}}_\mathcal{X}$ being conditionally independent of $\bm{\widetilde{A}}_\mathcal{X}$ and $\bm{A}$ given $\bm{B}$. This shows that the information leakage to any $X$ colluding workers is zero. Hence, the scheme is $X$\nobreakdash-secure.
\end{IEEEproof}

The next question is whether the MDS property of the codes $\mathcal{C}_A^\security$ and $\mathcal{C}_B^\security$ is needed for the security. If we did not require that the security property has to hold for all distributions of $\bm{A}$ and $\bm{B}$, then the MDS property would not be needed if $H(\bm{A}) = 0$ or $H(\bm{B}) = 0$, since there is no information to leak in the first place. The following lemma will show that under certain conditions, the codes need to be MDS.

\begin{lemma}\label{lem:dual_distance_security}
Let $d_A^\perp$ and $d_B^\perp$ be the minimum distances of $\mathcal{C}_A^\perp$ and $\mathcal{C}_B^\perp$. If $X \leq \min \{ d_A^\perp, d_B^\perp \} - 1$, then the linear SDMM scheme is $X$\nobreakdash-secure if and only if $\mathcal{C}_A^\security$ and $\mathcal{C}_B^\security$ are MDS codes.
\end{lemma}

\begin{IEEEproof}
If $\mathcal{C}_A^\security$ and $\mathcal{C}_B^\security$ are MDS codes, then the security is clear by Proposition~\ref{prop:linear_SDMM_security}. Hence, assume that the scheme is $X$\nobreakdash-secure. Let $\bm{A}$ be uniformly distributed and $\mathcal{X} \subseteq [N]$, $|\mathcal{X}| = X$, be a set of colluding workers. We have that any $d_A^\perp - 1$ columns of $F$ are linearly independent, so $\bm{\widetilde{A}}_\mathcal{X}$ is uniformly distributed. Therefore,
\begin{equation*}
    I(\bm{A}; \bm{\widetilde{A}}_\mathcal{X}) = H(\bm{\widetilde{A}}_\mathcal{X}) - H(\bm{R}'_\mathcal{X}) = 0
\end{equation*}
if and only if $H(\bm{R}'_\mathcal{X}) = H(\bm{\widetilde{A}}_\mathcal{X})$, \emph{i.e.}, if and only if $\bm{R}'_\mathcal{X}$ is uniformly distributed. Thus, $F^{> mp}_\mathcal{X}$ is invertible and $\mathcal{C}_A^\security$ is an MDS code. Similarly, we get that $\mathcal{C}_B^\security$ is MDS.
\end{IEEEproof}

The above lemma is useful when studying linear SDMM schemes constructed from MDS codes.

\begin{corollary}\label{cor:MDS_code_security}
If $\mathcal{C}_A$ and $\mathcal{C}_B$ are MDS codes, then the linear SDMM scheme is $X$\nobreakdash-secure if and only if $\mathcal{C}_A^\security$ and $\mathcal{C}_B^\security$ are MDS codes.
\end{corollary}

\begin{IEEEproof}
By properties of MDS codes, we get that $d_A^\perp = N - (N - (mp + X)) + 1 = mp + X + 1$, so $X \leq d_A^\perp - 1 = mp + X$. Similarly, $X \leq d_B^\perp - 1 = np + X$. The result follows from Lemma~\ref{lem:dual_distance_security}.
\end{IEEEproof}

\subsection{Bounds for Linear SDMM}\label{sec:linear_SDMM_bounds}

We will only consider linear SDMM schemes which are decodable and secure against $X$\nobreakdash-collusion. As an immediate consequence of Proposition~\ref{prop:product_singleton_bound} (Theorem 2 in \cite{randriambololona2013upper}) we get the following lower bound for the recovery threshold for a linear SDMM scheme.

\begin{theorem}\label{thm:recovery_threshold}
A linear SDMM scheme has recovery threshold
\begin{equation*}
    R \geq \min\{N, (m + n)p + 2X - 1\}.
\end{equation*}
\end{theorem}

\begin{IEEEproof}
We define $R = N - D + 1$, where $D$ is the minimum distance of the code $\mathcal{C}_A \star \mathcal{C}_B$. The codes $\mathcal{C}_A$ and $\mathcal{C}_B$ have length $N$ and dimensions $mp + X$ and $np + X$, respectively. Therefore,
\begin{equation*}
    D \leq \max \{ 1, N - (mp + X) - (np + X) + 2 \}
\end{equation*}
by Proposition~\ref{prop:product_singleton_bound}. Thus,
\begin{equation*}
    R = N - D + 1 \geq \min \{ N, (m + n)p + 2X - 1 \}. \IEEEQEDhereeqn
\end{equation*}
\end{IEEEproof}

We see that a linear SDMM scheme can achieve a recovery threshold lower than $(m + n)p + 2X - 1$ only when $R = N$ by the above theorem, \emph{i.e.}, when the scheme cannot tolerate stragglers. Therefore, we get the following theorem as a corollary.

\begin{theorem}\label{thm:recovery_threshold_stragglers}
A linear SDMM scheme that can tolerate stragglers has recovery threshold
\begin{equation*}
    R \geq (m + n)p + 2X - 1.
\end{equation*}
\end{theorem}

Another approach uses Proposition~\ref{prop:star_product_dimension_mds} (Theorem 7 in \cite{mirandola2015critical}) to find another lower bound for the recovery threshold. This theorem uses the natural security condition of Proposition~\ref{prop:linear_SDMM_security}.

\begin{theorem}\label{thm:mds_security_recovery_threshold}
A linear SDMM scheme with MDS codes $\mathcal{C}_A^\security$ and $\mathcal{C}_B^\security$ has recovery threshold
\begin{equation*}
    R \geq mn + \max \{m, n\}p + 2X - 1.
\end{equation*}
\end{theorem}

\begin{IEEEproof}
We can use the decomposition of the codes to write
\begin{align*}
    &\mathcal{C}_A \star \mathcal{C}_B = (\mathcal{C}_A^\encoding \oplus \mathcal{C}_A^\security) \star (\mathcal{C}_B^\encoding \oplus \mathcal{C}_B^\security) = \mathcal{C}_A^\encoding \star \mathcal{C}_B^\encoding + \mathcal{C}_A^\security \star \mathcal{C}_B^\encoding + \mathcal{C}_A^\encoding \star \mathcal{C}_B^\security + \mathcal{C}_A^\security \star \mathcal{C}_B^\security.
\end{align*}
Let us consider the linear decoding map given by
\begin{equation*}
    \widetilde{C} \mapsto \sum_{i \in [N]} \Lambda_i \otimes \widetilde{C}_i.
\end{equation*}
By writing $\widetilde{C} = (A' + R') \star (B' + S')$ we get
\begin{align*}
    AB = \sum_{i \in [N]} \Lambda_i \otimes \widetilde{C}_i = \sum_{i \in [N]} \Lambda_i \otimes A'_i B'_i + \sum_{i \in [N]} \Lambda_i \otimes \left( A'_i S'_i + R'_iB'_i + R'_iS'_i \right).
\end{align*}
As this has to hold for all choices of the random matrices, it has to hold when they are chosen to be zeros. Hence,
\begin{equation*}
    \sum_{i \in [N]} \Lambda_i \otimes \left( A'_i S'_i + R'_iB'_i + R'_iS'_i \right) = 0
\end{equation*}
for all choices of the random matrices. By picking out any entry of the response matrices, we get a linear map
\begin{equation*}
    \Dec \colon \mathcal{C}_A \star \mathcal{C}_B \to \F_q^{m \times n}.
\end{equation*}
By the rank--nullity theorem,
\begin{equation*}
    \dim \mathcal{C}_A \star \mathcal{C}_B = \dim\im(\Dec) + \dim\ker(\Dec).
\end{equation*}
From the previous computation and the decomposition of the codes, we see that
\begin{align*}
    &\mathcal{C}_A \star \mathcal{C}_B^\security + \mathcal{C}_A^\security \star \mathcal{C}_B^\encoding = \mathcal{C}_A^\security \star \mathcal{C}_B + \mathcal{C}_A^\encoding \star \mathcal{C}_B^\security \\
    &= \mathcal{C}_A^\security \star \mathcal{C}_B^\encoding + \mathcal{C}_A^\encoding \star \mathcal{C}_B^\security + \mathcal{C}_A^\security \star \mathcal{C}_B^\security \subseteq \ker \Dec.
\end{align*}
Using Proposition~\ref{prop:star_product_dimension_mds} we can give a lower bound on the dimension of $\ker \Dec$, since $\mathcal{C}_B^\security$ is MDS. Thus,
\begin{align*}
    \dim\ker(\Dec) \geq \dim\mathcal{C}_A \star \mathcal{C}_B^\security \geq \min \{ N, (mp + X) + X - 1 \}.
\end{align*}
The minimum cannot be $N$, since then $\dim\ker(\Dec) = N$, so $\Dec$ is the zero map. Hence, the minimum is achieved by the second term. On the other hand, the output space of $\Dec$ is $mn$ dimensional, since we must be able to produce any matrix. Combining this with the dimension of $\ker(\Dec)$ we get
\begin{equation*}
    \dim \mathcal{C}_A \star \mathcal{C}_B \geq mn + mp + 2X - 1.
\end{equation*}
Symmetrically, we get
\begin{equation*}
    \dim \mathcal{C}_A \star \mathcal{C}_B \geq mn + np + 2X - 1
\end{equation*}
by switching $m$ and $n$. These two inequalities give us the claimed inequality, since $R \geq \dim \mathcal{C}_A \star \mathcal{C}_B$.
\end{IEEEproof}

The above bound is well-known for GASP codes coming from the combinatorics of the degree table \cite[Theorem 2]{d2021degree}. The security of the GASP codes is proven by constructing the scheme such that $\mathcal{C}_A^\security$ and $\mathcal{C}_B^\security$ are MDS codes. Hence, we can see the above theorem as a generalization of this result. We notice that the bound on the recovery threshold given in Theorem~\ref{thm:mds_security_recovery_threshold} is quite loose in the case where $m, n, p > 1$ as seen in the construction in \cite{karpuk2023modular}. We do not believe that the bound in Theorem~\ref{thm:mds_security_recovery_threshold} is tight for all parameters.

\begin{remark}\label{rmk:optimal_schemes}
The SDMM scheme based on the DFT in \cite{mital2022secure} meets the bound in Theorem~\ref{thm:mds_security_recovery_threshold} since it has parameters $m = n = 1$ and $R = N = p + 2X$. Furthermore, the secure MatDot scheme in \cite{aliasgari2020private} meets the bound in Theorem~\ref{thm:recovery_threshold_stragglers} for linear SDMM schemes that can tolerate stragglers, since it has parameters $m = n = 1$ and $R = 2p + 2X - 1$. To the best of our knowledge, these optimality results have not been stated before. The linear SDMM framework is the first sufficiently general framework that has been studied and can be used to show optimality. It is still possible to have schemes that outperform the DFT or secure MatDot schemes, but these would have to be nonlinear or otherwise deviate from the given framework.
\end{remark}

Both Theorem~\ref{thm:recovery_threshold_stragglers} and Theorem~\ref{thm:mds_security_recovery_threshold} have the common term $2X$ in the bound, which gives that the number of colluding workers is strictly less than half of the number of workers.

\begin{corollary}\label{cor:number_of_colluding_workers}
A linear SDMM scheme with MDS codes $\mathcal{C}_A$ and $\mathcal{C}_B$ can tolerate at most $X < \frac{N}{2}$ colluding workers.
\end{corollary}

\begin{IEEEproof}
If $\mathcal{C}_A$ and $\mathcal{C}_B$ are MDS codes, then the bound given in Theorem~\ref{thm:mds_security_recovery_threshold} holds by Corollary~\ref{cor:MDS_code_security}. Therefore,
\begin{equation*}
    N \geq R \geq mn + \max\{m, n\}p + 2X - 1 \geq 2X + 1 > 2X
\end{equation*}
as $m, n, p \geq 1$.
\end{IEEEproof}

\subsection{Constructing SDMM Schemes Using the Framework}\label{sec:constructing_linear_SDMM_schemes}

The examples of SDMM schemes presented in Section~\ref{sec:examples_of_SDMM_schemes} can be described using the linear SDMM framework by describing the partitioning of the matrices, the codes $\mathcal{C}_A$ and $\mathcal{C}_B$, and the decoding process. Furthermore, the security of the schemes can be proven using Proposition~\ref{prop:linear_SDMM_security}.

\begin{example}[Secure MatDot]\label{ex:secure_matdot_linear_sdmm}
The secure MatDot scheme can be described using the linear SDMM framework as follows. The matrices $A \in \F_q^{t \times s}$ and $B \in \F_q^{s \times r}$ are partitioned into $p$ pieces using the inner product partitioning, \emph{i.e.}, $m = n = 1$ in the grid partitioning. The generator matrices $F$ and $G$ are defined as $(p + X) \times N$ Vandermonde matrices on the distinct evaluation points $\alpha_1, \dots, \alpha_N \in \F_q^\times$:
\begin{align*}
    F = \begin{pmatrix}
        1 & 1 & \cdots & 1 \\
        \alpha_1 & \alpha_2 & \cdots & \alpha_N \\
        \alpha_1^2 & \alpha_2^2 & \cdots & \alpha_N^2 \\
        \vdots & \vdots & \ddots & \vdots \\
        \alpha_1^{p + X - 1} & \alpha_2^{p + X - 1} & \cdots & \alpha_N^{p + X - 1}
    \end{pmatrix}, \quad
    G = \begin{pmatrix}
        \alpha_1^{p-1} & \alpha_2^{p-1} & \cdots & \alpha_N^{p-1} \\
        \vdots & \vdots & \ddots & \vdots \\
        \alpha_1 & \alpha_2 & \cdots & \alpha_N \\
        1 & 1 & \cdots & 1 \\
        \alpha_1^p & \alpha_2^p & \cdots & \alpha_N^p \\
        \vdots & \vdots & \ddots & \vdots \\
        \alpha_1^{p + X - 1} & \alpha_2^{p + X - 1} & \cdots & \alpha_N^{p + X - 1}
    \end{pmatrix}.
\end{align*}
These matrices generate Reed--Solomon codes of dimension $p + X$ and length $N$ on the evaluation points $\alpha = (\alpha_1, \dots, \alpha_N)$. We denote this by $\mathcal{C}_A = \RS_{p + X}(\alpha)$ and $\mathcal{C}_B = \RS_{p + X}(\alpha)$. It is easy to see that this produces the same encoding as the general description of the secure MatDot scheme. It was noted in \cite{mirandola2015critical} that the resulting star product code is then $\mathcal{C}_A \star \mathcal{C}_B = \RS_{2p + 2X - 1}(\alpha)$, provided that $N \geq 2p + 2X - 1$. The decoding can be done by computing
\begin{align*}
    \sum_{i \in [N]} [\lambda_i^{(p-1)}] \otimes \widetilde{C}_i &= \sum_{i \in [N]} \lambda_i^{(p-1)} \widetilde{C}_i \\
    &= \sum_{i \in [N]} \lambda_i^{(p-1)} h(\alpha_i) \\
    &= h^{(p-1)} = AB,
\end{align*}
where $\lambda_i^{(p-1)}$ is the coefficient of $x^{p-1}$ in the $i$th Lagrange interpolation polynomial on the evaluation points $\alpha$. Here $h(x)$ is the same product polynomial that is defined in Example~\ref{ex:secure_matdot} and $h^{(p-1)} = AB$ is the coefficient of $x^{p-1}$ in that polynomial. We have the decomposition
\begin{equation*}
    \mathcal{C}_A = \mathcal{C}_B = \RS_p(\alpha) \oplus \GRS_X(\alpha, \alpha^p),
\end{equation*}
where $\alpha^p = (\alpha_1^p, \dots, \alpha_N^p)$. Hence, the scheme is $X$\nobreakdash-secure by Proposition~\ref{prop:linear_SDMM_security} as $\GRS_X(\alpha, \alpha^p)$ is MDS. The recovery threshold of this scheme is $R = 2p + 2X - 1$, which meets the bound in Theorem~\ref{thm:recovery_threshold_stragglers}. Notice that the codes $\mathcal{C}_A$ and $\mathcal{C}_B$ are the same, but we use different generator matrices in the encoding phase. This shows that the choice of the generator matrices is important.
\end{example}

\begin{example}[GASP code]\label{ex:gasp_linear_sdmm}
We will continue Example~\ref{ex:gasp} to show how the GASP scheme can be described using linear SDMM. The matrices $A \in \F_q^{t \times s}$ and $B \in \F_q^{s \times r}$ are partitioned to $m = n = 3$ pieces using the outer product partitioning, \emph{i.e.}, $p = 1$ in the grid partitioning. The generator matrices are determined by the evaluation points $\alpha$ and the exponents in the polynomials $f(x)$ and $g(x)$. By choosing the same polynomials as in Example~\ref{ex:gasp} we get the generator matrices
\begin{align*}
    F = \begin{pmatrix}
        1 & 1 & \cdots & 1 \\
        \alpha_1 & \alpha_2 & \cdots & \alpha_N \\
        \alpha_1^2 & \alpha_2^2 & \cdots & \alpha_N^2 \\
        \alpha_1^9 & \alpha_2^9 & \cdots & \alpha_N^9 \\
        \alpha_1^{12} & \alpha_2^{12} & \cdots & \alpha_N^{12}
    \end{pmatrix}, \quad
    G = \begin{pmatrix}
        1 & 1 & \cdots & 1 \\
        \alpha_1^3 & \alpha_2^3 & \cdots & \alpha_N^3 \\
        \alpha_1^6 & \alpha_2^6 & \cdots & \alpha_N^6 \\
        \alpha_1^9 & \alpha_2^9 & \cdots & \alpha_N^9 \\
        \alpha_1^{10} & \alpha_2^{10} & \cdots & \alpha_N^{10}
    \end{pmatrix}.
\end{align*}
The star product of the codes $\mathcal{C}_A$ and $\mathcal{C}_B$ is generated by
\begin{equation*}
    H = \begin{pmatrix}
        1 & 1 & \cdots & 1 \\
        \alpha_1 & \alpha_2 & \cdots & \alpha_N \\
        \alpha_1^2 & \alpha_2^2 & \cdots & \alpha_N^2 \\
        \vdots & \vdots & \ddots & \vdots \\
        \alpha_1^{22} & \alpha_2^{22} & \cdots & \alpha_N^{22}
    \end{pmatrix},
\end{equation*}
where the exponents of the evaluation points are sums of the exponents of $f(x)$ and $g(x)$, \emph{i.e.},
\begin{equation*}
    \eta = (0, 1, 2, \dots, 12, 15, 18, 19, 21, 22).
\end{equation*}
By setting $N = 18$, we have that $H$ is an $18 \times 18$ matrix. The evaluation points $\alpha$ are chosen such that $H$ is invertible and that $\mathcal{C}_A^\security$ and $\mathcal{C}_B^\security$ are MDS codes. This can be done by utilizing the Schwartz--Zippel lemma over a large enough field. Thus, the scheme is $X$\nobreakdash-secure by Proposition~\ref{prop:linear_SDMM_security}.

We can reconstruct $AB$ by computing linear combinations of the responses. In particular, by setting
\begin{equation*}
    \Lambda_i = \begin{pmatrix}
        (H^{-1})_{i, 1} & (H^{-1})_{i, 4} & (H^{-1})_{i, 7} \\
        (H^{-1})_{i, 2} & (H^{-1})_{i, 5} & (H^{-1})_{i, 8} \\
        (H^{-1})_{i, 3} & (H^{-1})_{i, 6} & (H^{-1})_{i, 9}
    \end{pmatrix}
\end{equation*}
we can compute the linear combination
\begin{align*}
    \sum_{i \in [N]} \Lambda_i \otimes \widetilde{C}_i
    &= \sum_{i \in [N]} \begin{pmatrix}
        \widetilde{C}_i (H^{-1})_{i, 1} & \widetilde{C}_i (H^{-1})_{i, 4} & \widetilde{C}_i (H^{-1})_{i, 7} \\
        \widetilde{C}_i (H^{-1})_{i, 2} & \widetilde{C}_i (H^{-1})_{i, 5} & \widetilde{C}_i (H^{-1})_{i, 8} \\
        \widetilde{C}_i (H^{-1})_{i, 3} & \widetilde{C}_i (H^{-1})_{i, 6} & \widetilde{C}_i (H^{-1})_{i, 9}
    \end{pmatrix} \\
    &= \begin{pmatrix}
        A_1B_1 & A_1B_2 & A_1B_3 \\
        A_2B_1 & A_2B_2 & A_2B_3 \\
        A_3B_1 & A_3B_2 & A_3B_3
    \end{pmatrix} = AB.
\end{align*}
Here we utilize the equality
\begin{equation*}
    (A_1B_1, A_2B_1, \dots, A_3B_3, \dots) = (\widetilde{C}_1, \dots, \widetilde{C}_N)H^{-1}
\end{equation*}
which comes from the definition of the polynomial $h(x)$ in Example~\ref{ex:gasp}.
\end{example}

\begin{example}[SDMM based on DFT]\label{ex:sdmm_based_on_dft_linear_sdmm}
The SDMM scheme based on DFT that was first presented in \cite{mital2022secure} uses the inner product partitioning to partition the matrices to $p = N - 2X$ pieces. The generator matrices can be expressed as
\begin{align*}
    F = \begin{pmatrix}
    1 & 1 & \cdots & 1 \\
    1 & \zeta & \cdots & \zeta^{N - 1} \\
    1 & \zeta^2 & \cdots & \zeta^{2(N - 1)} \\
    \vdots & \vdots & \ddots & \vdots \\
    1 & \zeta^{p + X - 1} & \cdots & \zeta^{(p + X - 1)(N - 1)}
    \end{pmatrix}, \quad
    G = \begin{pmatrix}
    1 & 1 & \cdots & 1 \\
    1 & \zeta^{-1} & \cdots & \zeta^{-(N - 1)} \\
    \vdots & \vdots & \ddots & \vdots \\
    1 & \zeta^{-(p - 1)} & \cdots & \zeta^{-(p - 1)(N - 1)} \\
    1 & \zeta^{-(p + X)} & \cdots & \zeta^{-(p + X)(N - 1)} \\
    \vdots & \vdots & \ddots & \vdots \\
    1 & \zeta^{-(p + 2X - 1)} & \cdots & \zeta^{-(p + 2X - 1)(N - 1)}
    \end{pmatrix}.
\end{align*}
These follow directly from the general description in Example~\ref{ex:sdmm_based_on_dft}. From the generator matrices, we can see the decompositions
\begin{align*}
    \mathcal{C}_A &= \RS_p(\alpha) \oplus \GRS_X(\alpha, \alpha^p) \\
    &= \RS_{p+X}(\alpha) \\
    \mathcal{C}_B &= \RS_p(\alpha^{-1}) \oplus \GRS_X(\alpha^{-1}, \alpha^{-(p+X)}) \\
    &= \GRS_{p+X}(\alpha, \alpha^{-p}),
\end{align*}
where $\alpha = (1, \zeta, \zeta^2, \dots, \zeta^{N-1})$ and $\zeta$ is a primitive $N$th root of unity. Furthermore, $\alpha^k = (1, \zeta^k, \zeta^{2k}, \dots, \zeta^{k(N-1)})$. The star product of these codes is $\F_q^N$, so the recovery threshold is $R = N = p + 2X$, which is below the bound described in Theorem~\ref{thm:recovery_threshold_stragglers}. This is because the scheme is not able to tolerate stragglers. On the other hand, the scheme is able to reach the bound in Theorem~\ref{thm:mds_security_recovery_threshold}.
\end{example}

\begin{example}[Hermitian curve]\label{ex:hermitian_curve}
We shall consider an example coming from algebraic geometry codes. In particular, let us consider the Hermitian function field $H_2 = \F_4(x, y)$ defined by $y^2 + y = x^3$. By \cite[Lemma 6.4.4]{stichtenoth2009algebraic} this curve has genus $g = 1$ and $9$ rational places. Let $P_1, \dots, P_8, P_\infty$ be the rational places, where $P_\infty$ is the common pole of $x$ and $y$ and $P_1$ the zero of $y$, and define $\mathcal{P} = \{P_2, \dots, P_8\}$. Define the divisors $F = G = 3P_\infty$ and the length $N = 7$ algebraic geometry codes $\mathcal{C}_A = \mathcal{C}_\mathcal{L}(\mathcal{P}, F)$ and $\mathcal{C}_B = \mathcal{C}_\mathcal{L}(\mathcal{P}, G)$. The star product code is given by
\begin{equation*}
    \mathcal{C}_A \star \mathcal{C}_B = \mathcal{C}_\mathcal{L}(\mathcal{P}, F + G)
\end{equation*}
using \cite[Corollary 6]{couvreur2017cryptanalysis}, since $\deg F = \deg G = 3 \geq 2g + 1$. The generator matrices can be constructed by considering the Riemann--Roch spaces $\mathcal{L}(F)$ and $\mathcal{L}(G)$, which have bases $\{1, x, y\}$. Furthermore, $\mathcal{L}(F + G)$ has basis $\{1, x, y, x^2, xy, x^3\}$. By considering the defining equation, we may consider the basis $\{1, x, y, x^2, xy, y^2\}$, which is obtained as products of the bases of $\mathcal{L}(F)$ and $\mathcal{L}(G)$.

The matrices $A \in \F_4^{t \times s}$ and $B \in \F_4^{s \times r}$ are partitioned to $p = 2$ pieces using the inner product partitioning. We protect against $X = 1$ colluding workers. The generator matrices are defined as the generator matrices of $\mathcal{C}_A$ and $\mathcal{C}_B$ using the bases described above. Thus,
\begin{equation*}
    F = G = \begin{pmatrix}
        1 & 1 & \cdots & 1 \\
        x(P_2) & x(P_3) & \cdots & x(P_8) \\
        y(P_2) & y(P_3) & \cdots & y(P_8)
    \end{pmatrix}.
\end{equation*}
The encoded pieces are evaluations of $A_1 + A_2x + R_1y$ and $B_1 + B_2x + S_1y$ at the places $P_2, \dots, P_8$. Then we have that $A_1B_1$ is the coefficient of $1$ in the responses and $A_2B_2$ is the coefficient of $x^2$. Hence, the product $AB = A_1B_1 + A_2B_2$ can be computed as a linear combination of the responses. The resulting code $\mathcal{C}_A \star \mathcal{C}_B$ has minimum distance $D = 1$. Hence, the scheme has a recovery threshold $R = N - D + 1 = 7$. Furthermore, the scheme is $1$\nobreakdash-secure, since $\mathcal{C}_A^\security = \mathcal{C}_B^\security$ are full-support codes.

The secure MatDot scheme with the same parameters, $p = 2$ and $X = 1$, has a recovery threshold $2p + 2X - 1 = 5$ and can tolerate straggling workers. It seems nontrivial to construct a decodable and $X$\nobreakdash-secure linear SDMM scheme using algebraic geometry codes.
\end{example}

Algebraic geometry codes have recently been studied in SDMM with the HerA construction \cite{machado2023hera}, which is based on the Hermitian curve, as well as in \cite{makkonen2023algebraic} with the PoleGap construction, which is based on Kummer extensions. Both of these schemes fit in the linear SDMM framework as they choose $\mathcal{C}_A$ and $\mathcal{C}_B$ to be suitable AG codes.

Recently, constructions using grid partitioning have been given in the literature with general parameters $m, n, p > 1$. The Modular Polynomial scheme presented in \cite{karpuk2023modular} follows a similar linear structure that is given in the linear SDMM framework, where the matrix partitions are encoded using suitable linear codes.

\begin{remark}
Not all SDMM schemes from the literature can be described using the linear SDMM framework. The field trace polynomial code presented in \cite{machado2021field} uses a large field $\F_q$ while the responses are in some subfields of $\F_q$. This reduces the download cost since the elements of the smaller fields use less bandwidth. On the other hand, it is not possible to utilize this construction over prime fields that may be preferred in some applications. As the linear SDMM framework does not account for the different fields it is not possible to describe the field trace polynomial code using it. However, the linear structure is still present in the field trace polynomial code.
\end{remark}

\section{Error Correction in SDMM}\label{sec:error_correction_SDMM}

Protecting against straggling workers has been the subject of research in many SDMM schemes. Another form of robustness is protection against so-called Byzantine workers, which return erroneous responses as a result of a fault or on purpose. This error can occur during the computation or transmission, but we assume that the number of errors is bounded below parameter $E$. Robustness against Byzantine workers has been studied in the context of private information retrieval (PIR) and other distributed computation systems such as Lagrange coded computation in \cite{yu2019lagrange}.

The difference between straggling workers and Byzantine workers is that a straggling worker is simple to detect while noticing erroneous responses from a Byzantine worker is not as straightforward. In coding-theoretic terms, the straggling workers correspond to erasures in codes and Byzantine workers correspond to errors. It is well-known that erasures require one additional code symbol to fix with MDS codes, while errors typically require two additional code symbols to fix. The authors of \cite{yu2019lagrange} devised a coded computation scheme, where each additional straggler requires one additional response and each Byzantine worker requires two additional responses. This disparity between the costs can be fixed using interleaved codes by utilizing the structure of the error patterns.

\subsection{Interleaved Codes in SDMM}\label{sec:interleaved_codes_SDMM}

\begin{figure}[!t]
    \centering
    \includegraphics[width=0.3\columnwidth]{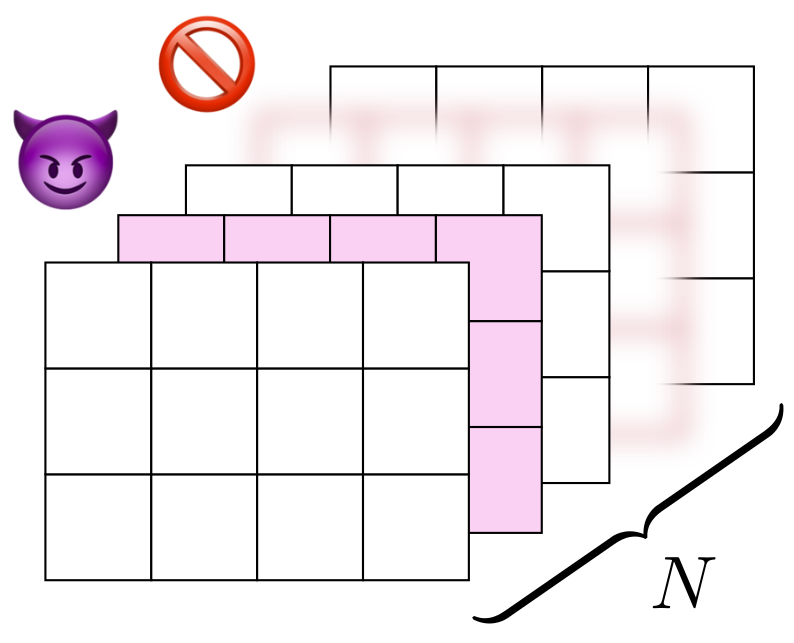}
    \caption{Diagram depicting the responses from the worker nodes. The Byzantine worker is depicted by the purple layer and the straggler by the blurred layer. Each response is a matrix, which is represented as a rectangular array in the figure. The codewords are the length $N$ vectors formed by stacking the responses and looking at the corresponding matrix entries. Hence, a Byzantine worker and stragglers can only affect their own position in the codewords.}
    \label{fig:response_diagram}
\end{figure}

The responses of the workers in a linear SDMM scheme can be expressed as $\widetilde{C}_i + Z_i$, where $Z_i$ is a potentially nonzero error matrix and $\widetilde{C}_i = \widetilde{A}_i\widetilde{B}_i$. We require that the number of (nonzero) errors is at most $E$, \emph{i.e.}, there are at most $E$ Byzantine workers. We may consider each of the individual codewords of the matrix code by considering a specific matrix entry, say $(\alpha, \gamma)$, of the responses. Such a vector is of the form
\begin{equation*}
    \widetilde{C}^{\alpha\gamma} + Z^{\alpha\gamma} \in \F_q^N,
\end{equation*}
where $\widetilde{C}^{\alpha\gamma} \in \mathcal{C}_A \star \mathcal{C}_B$. As $\wt(Z^{\alpha\gamma}) \leq E$, we may uniquely correct the errors if $D \geq 2E + 1$, where $D$ is the minimum distance of $\mathcal{C}_A \star \mathcal{C}_B$. Additionally, if there are $S$ stragglers, then we need $D \geq 2E + S + 1$, which corresponds to the well-known bound for bounded distance decoding.

Let $\mathcal{E} \subseteq [N]$ be the indices of the Byzantine workers. Then $\supp Z^{\alpha\gamma} \subseteq \mathcal{E}$ for all matrix positions $(\alpha, \gamma)$, which means that the errors are located in the same places in all codewords. This corresponds to burst errors in the associated interleaved code. There are several algorithms for decoding interleaved codes that can correct up to twice as many errors as non-interleaved decoders, such as those presented in \cite{schmidt2009collaborative, holzbaur2021success}. This is achieved by collaborative decoding, where the fact that the erroneous symbols are in the same place in each codeword is utilized.

Figure \ref{fig:response_diagram} depicts how the responses of a linear SDMM scheme can be seen as a collection of codewords from the star product code $\mathcal{C}_A \star \mathcal{C}_B$. Each layer in the diagram depicts the responses from one of the workers. By collecting the matching matrix entries to a vector of length $N$ we obtain codewords in the code $\mathcal{C}_A \star \mathcal{C}_B$ with some possible errors. If one of the workers returns an incorrect result, say worker 2 in Figure \ref{fig:response_diagram}, then the errors in the codewords will be in coordinate 2. Similarly, if one of the workers fails to return a response in time, say worker 4 in Figure \ref{fig:response_diagram}, then the corresponding coordinate is an erasure in each of the codewords.

Our proposed idea for correcting errors from the responses of a linear SDMM scheme with at most $E$ Byzantine workers is the following.
\begin{itemize}
    \item Compute the syndromes of each of the vectors in the response matrices and find out which matrix entries contain errors.
    \item Choose some subset of $\ell$ matrix entries which contain errors and collect the corresponding $\ell$ vectors as a codeword of the $\ell$\nobreakdash-interleaved code.
    \item Find the error locations from the interleaved code using an error correction algorithm for the $\ell$\nobreakdash-interleaved code.
    \item Treat the erroneous coordinates as erasures and decode as usual.
\end{itemize}
As error correction of the interleaved codewords requires more computation compared to decoding without errors, it is not advantageous to choose $\ell$ to be maximal, \emph{i.e.}, choosing all of the matrix entries to the interleaved codeword. On the other hand, collaborative decoding algorithms do not guarantee success with probability 1, so $\ell$ has to be chosen such that the success probability is suitably high.

\subsection{Analyzing Error Correction Capabilities}\label{sec:analyzing_error_correction}

Interleaved coding techniques can be used with any linear SDMM scheme. However, many codes that are used in different SDMM constructions do not have efficient error correction algorithms. SDMM schemes that are based on polynomial interpolation, such as the secure MatDot or $\mathrm{GASP}_\text{big}$ schemes, can be utilized, since Reed--Solomon codes have well-known error correction algorithms. Collaborative error correction algorithms have been designed for interleaved Reed--Solomon codes since they are prevalent in many applications where burst errors are common. In this section, we analyze the success probability of some interleaved Reed--Solomon decoders in the context of the secure MatDot and $\mathrm{GASP}_\text{big}$ schemes. The same techniques are applicable to other linear SDMM schemes based on Reed--Solomon codes.

We assume that the errors sent by the Byzantine workers are uniformly distributed, \emph{i.e.}, the errors $Z_i$ for $i \in \mathcal{E}$ are independent and uniformly distributed. This is a natural assumption if the errors occur naturally without malice. Additionally, this assumption is popular in the literature, where failure probabilities are analyzed.

Bounded distance decoders for interleaved Reed--Solomon codes are discussed in \cite{schmidt2009collaborative, holzbaur2021success}. These decoding algorithms generalize the Berlekamp--Massey approach of decoding Reed--Solomon codes to interleaved codes. Additionally, \cite{schmidt2009collaborative, holzbaur2021success} give bounds on the success probability of the decoders when the errors are assumed to be uniformly distributed with specified column weights.

\begin{theorem}\label{thm:failure_probability}
Consider a linear SDMM scheme over $\F_q$ where $\mathcal{C}_A \star \mathcal{C}_B$ is a Reed--Solomon code with minimum distance $D$. If there are at most $D - 2$ Byzantine workers, which return independent and uniform errors, then there is an error correction algorithm, which will correct the errors with failure probability at most
\begin{equation*}
    \left( \frac{q^\ell - q^{-1}}{q^\ell - 1} \right)^{D - 2} \cdot \frac{q^{D - 2 - \ell}}{q - 1},
\end{equation*}
where $\ell$ is the chosen interleaving order.
\end{theorem}

\begin{IEEEproof}
As concluded in the discussion above, the errors caused by the Byzantine workers are burst errors in the $\ell$\nobreakdash-interleaved Reed--Solomon code. Furthermore, the errors are distributed uniformly by assumption. Therefore, we can utilize \cite[Theorem 7]{holzbaur2021success}, which states that the probability of unsuccessful decoding is at most
\begin{equation*}
    \left( \frac{q^\ell - q^{-1}}{q^\ell - 1} \right)^t \cdot \frac{q^{-(\ell + 1)(t_\text{max} - t)}}{q - 1},
\end{equation*}
where $t$ is the number of errors and $t_\text{max} = \frac{\ell}{\ell + 1}(D - 1)$. As $t \leq D - 2$ by  assumption, we get that the probability of unsuccessful decoding is at most
\begin{align*}
    &\quad\left( \frac{q^\ell - q^{-1}}{q^\ell - 1} \right)^{D - 2} \cdot \frac{q^{-(\ell(D + 1) - (\ell + 1)(D - 2))}}{q - 1} \\
    &= \left( \frac{q^\ell - q^{-1}}{q^\ell - 1} \right)^{D - 2} \cdot \frac{q^{D - 2 - \ell}}{q - 1}
\end{align*}
since the expression is increasing in $t$.
\end{IEEEproof}

We assume that the field size $q$ is suitably large since this is natural in settings where the matrices are discretized from the real numbers or the integers. The field size would be of the order of $2^{32}$ or $2^{64}$ to make implementation efficient.

We may now choose a suitable interleaving order $\ell$ to make the probability of unsuccessful decoding suitably low. We see that for large $q$, the upper bound given in Theorem~\ref{thm:failure_probability} is approximately $q^{D - 3 - \ell}$, since the first term is approximately 1. Thus, for $\ell \geq D - 2$ we have that the probability of unsuccessful decoding is strikingly small. Choosing a larger $\ell$ will yield even lower failure probabilities. However, a larger interleaving order will naturally incur more computation in the collaborative decoding phase. Hence, we get a trade-off between the probability of unsuccessful decoding and the computational complexity.

With the assumption of Theorem~\ref{thm:failure_probability}, \emph{i.e.}, that the error matrices from the Byzantine workers are uniformly distributed, we see that we can correct up to $E = D - 2$ errors with high probability. Hence, we need a total of $N = R + S + E + 1$ workers to account for the $S$ straggling workers and $E$ Byzantine workers. This is an improvement over independent decoding of the codewords in the response matrices, which requires $N = R + S + 2E$ workers.

\subsection{Randomized Linear SDMM}

In the previous analysis, we assumed that the Byzantine workers return errors that are uniformly and independently distributed. This is a natural assumption if the errors occur during communication. However, the Byzantine workers may be able to introduce errors from other distributions or by specifically designing them such that the probability of unsuccessful decoding is much higher than what is indicated by Theorem~\ref{thm:failure_probability}.

Our proposed method is based on randomization of the linear SDMM scheme. In particular, we present a randomized secure MatDot scheme, which will make it more difficult for the Byzantine workers to craft malicious responses that cannot be corrected by the collaborative decoding method.

The randomized secure MatDot scheme is based on the secure MatDot scheme. Let $\widetilde{A}_i^\text{MatDot}$ and $\widetilde{B}_i^\text{MatDot}$ be the encoded matrices sent to the $i$th worker in the secure MatDot scheme. Furthermore, let $U_i$ and $V_i$ be random invertible diagonal matrices of suitable size chosen uniformly at random over $\F_q$. The worker is sent
\begin{equation*}
    \widetilde{A}_i^\text{rand} = U_i^{-1} \widetilde{A}_i^\text{MatDot}, \quad \widetilde{B}_i^\text{rand} = \widetilde{B}_i^\text{MatDot} V_i^{-1}.
\end{equation*}
This does not increase the computational complexity of the user, since multiplication by a diagonal matrix is proportional to the size of the matrix. The responses of the workers are of the form
\begin{equation*}
    \widetilde{A}_i^\text{rand}\widetilde{B}_i^\text{rand} + Z_i = U_i^{-1} \widetilde{A}_i^\text{MatDot}\widetilde{B}_i^\text{MatDot} V_i^{-1} + Z_i,
\end{equation*}
where $Z_i$ is a potentially nonzero error matrix. By multiplying this with $U_i$ and $V_i$ we obtain the responses
\begin{equation*}
    \widetilde{A}_i^\text{MatDot}\widetilde{B}_i^\text{MatDot} + U_iZ_iV_i.
\end{equation*}
These are responses in the secure MatDot scheme, but the errors are now of the form $U_iZ_iV_i$, where $U_i, V_i$ are random invertible diagonal matrices. Hence, we may use the error correction method highlighted in the previous section to correct the error. We call this scheme the \emph{randomized secure MatDot scheme}, since we essentially use randomized generalized Reed--Solomon codes in the encoding phase.

As the workers do not know the matrices $U_i$ and $V_i$, it is more difficult for them to coordinate the error matrix in a way that is favorable to them. The hope is that the Byzantine workers would return uniform errors, which means that the bound given in Theorem~\ref{thm:failure_probability} is valid since $U_iZ_iV_i$ is uniformly distributed if $Z_i$ is uniformly distributed.

\subsection{Comparison to the Error Detection Method}

The system model in the SDMM schemes differs from the classical setup in coding theory, where a message is sent over an unreliable channel from a sender to a receiver. In SDMM schemes, the user has all the information necessary to compute the responses $\widetilde{A}_i \widetilde{B}_i$ of the workers. This knowledge can be used to detect Byzantine workers using Freivalds' algorithm \cite{freivalds1979fast}, which is a probabilistic algorithm to detect errors in the matrix multiplication $\widetilde{C}_i = \widetilde{A}_i \widetilde{B}_i$. The algorithm consists of choosing a random vector $x$ and computing the matrix-vector products $\widetilde{B}_ix$, $\widetilde{A}_i (\widetilde{B}_ix) = \widetilde{C}_ix$ and $(\widetilde{C}_i + Z_i)x$, and comparing the last two products. If these are different, then the error matrix $Z_i$ from the $i$th worker is nonzero, \emph{i.e.}, the $i$th worker is a Byzantine worker and should be ignored. It may still be the case that $Z_ix = 0$ even if $Z_i \neq 0$, but we can bound the probability of this happening if $x$ is chosen at random. This approach was successfully utilized in SDMM in \cite{hofmeister2022secure} and \cite{tang2022adaptive}. This error detection method requires three matrix-vector multiplications for a total of $\mathcal{O}(\frac{sr}{pn} + \frac{ts}{mp} + \frac{tr}{mn})$ operations.

On the other hand, the complexity of the interleaved decoder does not depend on the middle dimension $s$ as it only works on the $N$ received matrices of dimension $\frac{t}{m} \times \frac{r}{n}$. Furthermore, the interleaved decoder does not need the original matrices $A$ and $B$ as input, which makes it possible to use in scenarios where the matrices do not originate at the user. Such a system model has been considered in \cite{jia2021capacity}.

\section{Conclusions and Future Work}

In this paper, we introduced the linear SDMM framework, which can be used to study most of the SDMM schemes in the literature. This framework is based on coding theory and it works for all linear codes. This is in contrast to earlier works, which are heavily based on evaluation codes. Utilizing the generality of the framework,  we provided some first results deriving from known results for star product codes. As many SDMM schemes from the literature can be considered as special cases of the linear SDMM framework, the framework provides a simpler way to compare different SDMM schemes. Additionally, we studied Byzantine workers in the context of SDMM and introduced a way to utilize interleaved codes to correct a larger number of errors with high probability.

In Theorem~\ref{thm:recovery_threshold_stragglers} and Theorem~\ref{thm:mds_security_recovery_threshold} we give bounds for the recovery threshold and notice that in some special cases, there are linear SDMM schemes achieving these bounds. In general, we do not believe that these bounds are tight for arbitrary partitioning parameters. In the future, we would like to give sharper bounds or find schemes achieving the current bounds, and use these bounds to study the rate and capacity of linear SDMM schemes. Additionally, we would like to extend our framework to cover the use of field extensions and array codes. Finally, we would like to study how well the randomized secure MatDot scheme works in the presence of different error distributions.

\section*{Acknowledgment}
\addcontentsline{toc}{section}{Acknowledgment}

The authors would like to thank Dr.\ Elif Saçıkara for useful discussions about algebraic geometry codes and for providing Example~\ref{ex:hermitian_curve}.

\bibliographystyle{IEEEtran}
\bibliography{TIT}

\end{document}